\documentclass[
preprint,
amssymb,
amsmath
]{revtex4-1}

\usepackage{graphicx}
\usepackage{dcolumn}
\usepackage{bm}
\usepackage{bm, color}

\begin{document}

\title{Linear response in topological materials}

\author{Jonathan Noky}
\author{Yan Sun}
\email{ysun@cpfs.mpg.de}
\affiliation{Max Planck Institute for Chemical Physics of Solids, 01187 Dresden, Germany}

\begin{abstract}
The discovery of topological insulators and semimetals has
opened up a new perspective to understand materials. Owing to
the special band structure and enlarged Berry curvature,
the linear responses are strongly enhanced in topological materials.
The interplay of topological band structure and symmetries plays
a crucial role for designing new materials with strong and exotic new
electromagnetic responses and provides promising mechanisms
and new materials for the next generation of technological applications.
We \textcolor{black}{review} the fundamental concept of linear responses in
topological materials from the symmetry point of view
and discuss their potential applications.
\end{abstract}

\maketitle

\section{Background and Introduction}
After the discovery of topological insulators (TIs)~\cite{Kane2005,Kane2005b,bernevig2006quantum,konig2007quantum,fu2007topological,hsieh2008topological,Hasan2010ku,Qi2011RMP}, 
topological band theory
was successfully generalized into condensed matter\cite{chiu2016classification,po2017symmetry,song2018quantitative,song2017diagnosis}. In the last decade, the
combination of topological band theory and symmetry in solids has led to different
types of quantum topological states from insulators to 
semimetals~\cite{kitaev2009periodic,fu2011topological,Wan2011,Burkov2011de,Wang2012,hsieh2012topological,Young2012,ando2015topological,Weng2015,huang2015weyl,
Xu2015TaAs,Lv2015TaAs,bradlyn2016beyond,belopolski2016discovery,changxu2016,Change1600295,bradlyn2017topological,PhysRevLett.119.196403,PhysRevB.96.041201,Xue1603266,armitage2018weyl,schindler2018higher,
zhang2019catalogue,tang2019comprehensive,vergniory2019complete,zhengjia2019}.
Depending on the topological charges and symmetries, different exotic topological
surface states and bulk transport properties were theoretically predicted and
experimentally observed. Owing to these attractive properties, a lot of effort has
been devoted to the study of topological materials for the next generation of
technological applications.

The interplay between electromagnetic response theory
and symmetry breaking is another crucial part for the 
understanding of transport properties in quantum materials. 
Materials with strong or quantum electromagnetic response have
an extensive impact on the development of data storage,
information processing,   energy conversion, etc.
Topological band structures characterized by band inversion,
linear band crossings, and band anti-crossings with different
topological charges can strongly enhance the electromagnetic
responses, and even lead to quantized response
effects in some special situations. Since the perturbation from
external fields can change the symmetry of a system, 
the electromagnetic response effects also offer an effective way
to manipulate the topological states and, in turn, to control
the information processing.

The electromagnetic response can be understood from the perturbation 
approximation with different orders of the perturbation series. In electrical
transport, the higher order responses are normally much smaller
than the linear response term. Therefore, most of the electromagnetic responses can be
understood in the linear response region, especially the different types of 
the Hall effect in topological materials. However, in some special situations, the linear response term is forbidden by the symmetry of the system. Consequently, the
higher order response effect will dominate in these cases, although the signal is normally very
weak and not easy to detect experimentally. In this paper, we   give a
review of electromagnetic responses in topological materials
in the linear response region.

\section{Linear Response Theory}
In the linear response approximation, the change of an
observable $\hat{A}$ with respect to an external
field $\hat{F}=\hat{B}F$ can be evaluated via a linear response
as~\cite{crepieux2001,nolting2008}
\begin{equation}
\begin{aligned}
\delta A_{i}=\chi_{ij}^{A}F_{j}
\end{aligned}
\label{response}
\end{equation}
with the linear response tensor

\begin{equation}
\begin{aligned}
\chi_{ij}^{A}=\lim_{\varepsilon\to 0}\sum_{m,n}\frac{f(E_{m})-f(E_{n})}{E_{m}-E_{n}-i\varepsilon}\langle n|\hat{A_{i}}|m\rangle\langle m|\hat{B_{j}}|n\rangle,
\end{aligned}
\label{tensor}
\end{equation}
where $n$ and $m$ are the band indices. In a periodic system, the
eigenstates and eigenvalues are labeled by band index and
momentum $k$ as $|n,k\rangle$ and $E_{n,k}$, respectively.

In electrical transport, we are interested in the linear response
with respect to an external electric field, which means $\hat{F}=eE\hat{r}$.
By using the commutation relation $\hat{v}=\frac{i}{\hbar}[\hat{H},\hat{r}]$, one can transform the position
operator to the velocity operator and the position operator matrix
can be replaced by

\begin{equation}
\begin{aligned}
\langle n,k|\hat{r}|m,k\rangle=-i\hbar\frac{\langle n,k|\hat{v}|m,k\rangle}{E_{n,k}-E_{m,k}}.
\end{aligned}
\label{v-r}
\end{equation}

With this replacement, one arrives at the commonly used form of the linear response tensor
\begin{equation}
\begin{aligned}
	\chi_{ij}^{A}=i\hbar e\lim_{\varepsilon\to 0}\sum_{k}\sum_{m,n}\frac{f(E_{n,k})-f(E_{m,k})}{(E_{n,k}-E_{m,k}-i\varepsilon)(E_{n,k}-E_{m,k})}\langle n,k|\hat{A_{i}}|m,k\rangle\langle m,k|\hat{v_{j}}|n,k\rangle.
\end{aligned}
\label{tensor-2}
\end{equation}

The linear response tensor can be further separated into  
inter-band and intra-band contributions. This separation is 
essential for the study of topological materials. 

For the inter-band part, because $\varepsilon\to 0$, one can 
easily get 
\begin{equation}
\begin{aligned}
        \chi_{ij}^{A-I}=i\hbar e\sum_{k}\sum_{m\neq n}\frac{f(E_{n,k})-f(E_{m,k})}{(E_{n,k}-E_{m,k})^{2}}\langle n,k|\hat{A_{i}}|m,k\rangle\langle m,k|\hat{v_{j}}|n,k\rangle.
\end{aligned}
\label{tensor-inter-band}
\end{equation}

For the intra-band contribution, it becomes the Boltzmann formula with the
constant relaxation time approximation ($\varepsilon=\hbar/\tau$) ~\cite{ashcroft1976}
\begin{equation}
\begin{aligned}
\chi_{ij}^{A-II}=-e\tau\sum_{k}\sum_{n}\delta(E_{F}-E_{n,k})\langle n,k|\hat{A_{i}}|\textcolor{black}{n},k\rangle\langle \textcolor{black}{n},k|\hat{v_{j}}|n,k\rangle.
\end{aligned}
\label{tensor-intra-band}
\end{equation}

The relaxation time for the intra-band contribution is hard
to estimate correctly, therefore it is not easy to give an accurate 
theoretical evaluation. Nevertheless, one can use it to estimate the response 
qualitatively, which is especially important for the symmetry analysis.  
In contrast, the inter-band part is easier to deal with in theory and 
numerical calculations. It can be accurately predicted in real materials
as long as the electronic band structure is accurate enough. In particular, 
when we are interested in electrical current, the intrinsic part is just 
the Berry phase of the electronic band structure. Hence, the inter-band 
part plays a  crucial role in topological materials.

\section{Anomalous Hall Effect in Ferromagnetic Topological Materials}
Considering the observable $\hat{A}$ as electrical current density
$\hat{A}=-e\hat{v}/V$, the inter-band part of linear response tensor
for each separated $k-$point becomes the so-called Berry curvature ~\cite{Xiao2010,nagaosa2010}
\begin{equation}
\begin{aligned}
\Omega^{k}(n,k)=\sum_{m\neq n}\frac{\langle n,k|\hat{v_{i}}|m,k\rangle\langle m,k|\hat{v_{j}}|n,k\rangle}{(E_{n,k}-E_{m,k})^{2}}
\end{aligned}
\label{Berry-curvature}
\end{equation}
and the anomalous Hall conductivity (AHC) is dependent on the integral of the
Berry curvature over the whole Brillouin zone (BZ). Because the Berry curvature
is odd with respect to time reversal symmetry, the
anomalous Hall effect (AHE) can only exist in magnetic systems in the linear
response approximation.

The AHE was first observed in 1879 by E. H. Hall 
in two-dimensional ferromagnetic systems~\cite{hall1879}.
According to the current understanding, there are mainly two contributions
for the AHE: one is the extrinsic contribution from
impurity scattering and the other   is the intrinsic contribution 
from Berry phase effects of the electronic band 
structure of the ideal single crystal~\cite{karplus1954,chang1996,haldane2004,Xiao2010,nagaosa2010}. \textcolor{black}{Because~both symmetry and topology are tied to the ideal crystal, it is not possible to discuss the the extrinsic contribution from the topological point of view. In the following, we   therefore focus on the intrinsic~part.}

In topological materials, owing to the special band structure, the 
local Berry curvature is strongly enhanced, and the AHE is dominated 
by the intrinsic contribution. The~extreme example is the 
quantum anomalous Hall effect (QAHE), where only intrinsic contributions
exist, with a quantized transverse anomalous Hall conductance
in quanta of $e^{2}/h$ and zero longitudinal resistance~\cite{Haldane1988}.
The~topological charge in the QAHE is indexed by a topological invariant, the Chern number, which is given by the integral of the Berry 
curvature over the whole
2D BZ.In contrast to the quantum Hall effect, which is decided by
the external magnetic field~\cite{Halperin1982}, the QAHE originates from
the spin--orbit coupling (SOC) and intrinsic magnetism.
The~QAHE was theoretically proposed by F. D. M. Haldane  in 1988
by a model defined on a honeycomb lattice~\cite{Haldane1988}. Owing to the development
of topological band theory and thin film growth techniques,
the first QAHE was experimentally observed in chromium-doped
(Bi,Sb)$_2$Te$_3$~\cite{Yu2010,Chang2013}.

An effective model for the QAHE can be described by a $2\times2$
nodal line band structure~\cite{liu2016}
\begin{equation}
\begin{aligned}
H_{QAHE}=A(k_{x}\sigma_{x}+k_{y}\sigma_{y})+M(k)\sigma_{z},
\end{aligned}
\label{QAHEeq}
\end{equation}
where $\sigma_{i}$ are the Pauli matrices and 
$M(k)=M_{0}+M_{1}(k_{x}^{2}+k_{y}^{2})$.

\begin{figure}[h]
\centering
\includegraphics[width=0.90\textwidth]{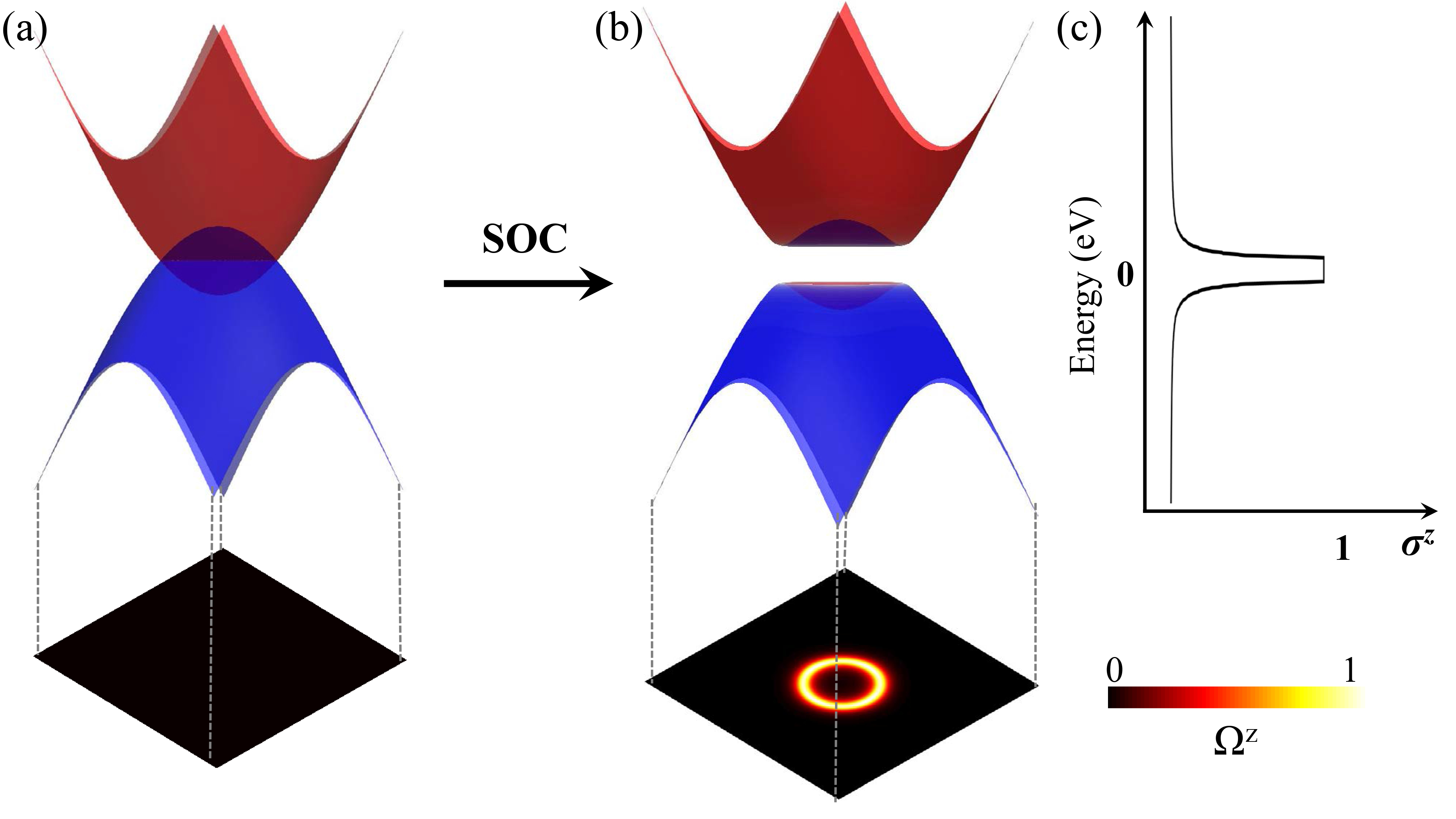}
\caption{
        Schematic of a quantum anomalous Hall insulator (QAHI) with a nodal line
        band structure.
        (\textbf{a}) The~band inversion forms a nodal ring in the situation without
        including SOC. The~Berry curvature is zero in the whole $k$ space.
        (\textbf{b}) The~nodal line is broken by SOC with opening a band gap, and
        non-zero Berry curvature forms a hot ring located in the band gap.
	The color bar for the Berry curvature is in
        arbitrary units. (\textbf{c}) A quantized anomalous Hall conductance ($e^2/h$)
        in the band gap.
}
\label{QAHEfig}
\end{figure}

Because this model includes a mirror symmetry with respect to $x-y$
plane, the band structure forms a nodal loop in the absence of
SOC, as presented in Figure~\ref{QAHEfig}a. 
This nodal line is broken by SOC and a band gap opens, which
generates non-zero Berry curvature in the band gap and forms a hot loop
in 2D $k$ space  (see Figure~\ref{QAHEfig}b). The~integral of the Berry curvature in the
whole 2D BZ leads to a quantized anomalous Hall conductance in the band gap,
which is just corresponding to the chiral edge state, 
as presented in Figure~\ref{QAHEfig}c.

\begin{figure}[h]
\centering
\includegraphics[width=0.90\textwidth]{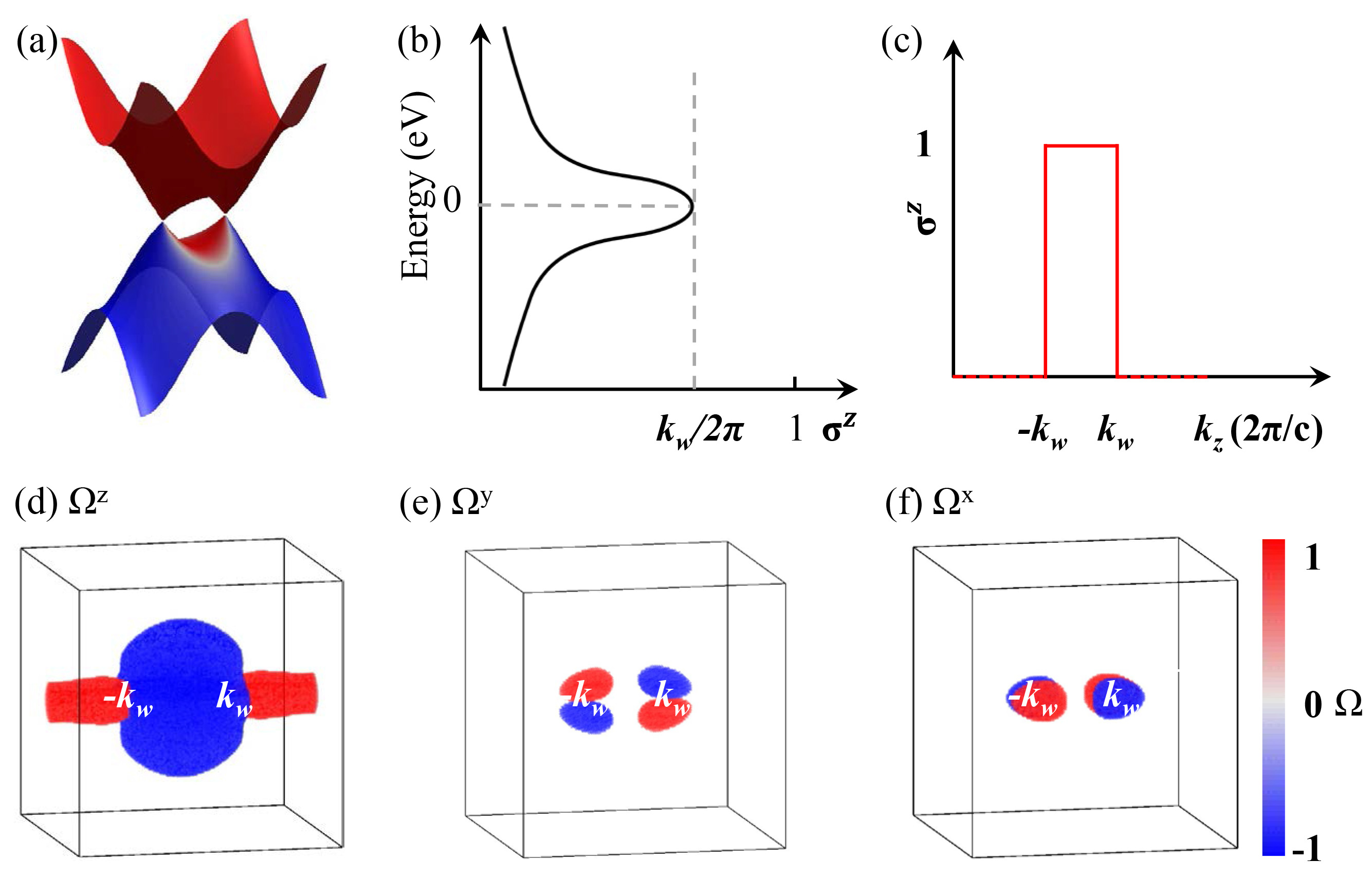}
\caption{
        AHE in magnetic WSMs.
        (\textbf{a}) Energy dispersion of a magnetic WSM with one 
	pair of Weyl points
        lying on the $z$-axis and at zero energy.
        (\textbf{b}) Energy dependent AHC for the effective model of a magnetic WSM.
        (\textbf{c}) Evolution of the anomalous Hall conductance for 
	each $k_x-k_y$ plane with varying
        $k_z$.
        (\textbf{d}--\textbf{g}) Berry curvature distribution in 3D BZ for the three 
	components of $\Omega^z$, $\Omega^y$, and $\Omega^x$. 
	The hot spots are mainly focused around the Weyl points. 
        The~color bar is in arbitrary units.
}
\label{Weyl}
\end{figure}

A stack of 2D QAHE systems was proposed to be an effective way to 
obtain magnetic Weyl semimetals (WSMs) with the
QAHE in its 3D version~\cite{burkov2011,zyuzin2012,Luhaizhou2015}. 
The simplest model can be obtained by including 
the coupling effect along $z$ via adding a $k_z$ term 
$M_{1}k_{z}^{2}\sigma_z$ into Equation \eqref{QAHEeq}~\cite{Luhaizhou2015}, 
where one pair of Weyl points is located on the $k_z$-axis at $(0,0,\pm k_w)$ with $k_w=\sqrt{M_{0}/M_{1}}$, 
as presented in Figure~\ref{Weyl}a.

The peak value of the anomalous Hall conductivity (AHC) lies at the energy of
the pair of Weyl points (Figure~\ref{Weyl}b). The~AHC in magnetic Weyl semimetals can 
be understood from 
the QAHE. The~Weyl points act as a sink and a source of the 
Berry curvature, just like monopoles in reciprocal space, 
therefore the $k_x-k_y$ planes host a non-zero quantized 
anomalous Hall conductance when $k_z$ is lying between the pair of 
Weyl points, and the contribution is 
zero for the planes with $k_{z}>|\sqrt{M_{0}/M_{1}}|$~\cite{Wan2011,Burkov2011de}. 
As a result, the AHC for the magnetic 
WSMs with one pair of Weyl point is 
$\sigma^{z}=\frac{e^{2}}{h}\frac{k_{w}}{\pi}$
and its magnitude is decided by the separation of the Weyl points in $k$ space. 
Because the time reversal symmetry is broken due to the magnetization along $z$,
only the $z$-component   of the Berry curvature shows a non-zero value after the 
integral over the whole $k$ space   (see Figure~\ref{Weyl}d). In contrast, the other two 
components still follow time reversal symmetry, as shown in Figure~\ref{Weyl}e,f,
and therefore they do not contribute to the~AHE.

\unskip

Because the nodal lines and Weyl points can generate strong local Berry curvature,
magnetic nodal line semimetals and WSMs are expected to have a strong intrinsic
AHE. Recently, a large AHE was observed in the
Heusler compound Co$_2$MnGa~\cite{kubler2012,manna2018} and the quasi-two-dimensional compounds Co$_3$Sn$_2$S$_2$~\cite{wang2018,liu2018}
and Fe$_3$GeTe$_2$~\cite{kim2018}, 
where the mirror symmetry protected
nodal line band structure is the most important feature for this phenomenon.
Particularly, owing to the topological band structures
and low charge carrier density, Co$_3$Sn$_2$S$_2$ and Co$_2$MnGa
are so far the only two examples of all known materials hosting both giant AHC and anomalous Hall angle (AHA)~\cite{wang2018,liu2018,manna2018}.

\begin{figure}[h]
\centering
\includegraphics[width=0.90\textwidth]{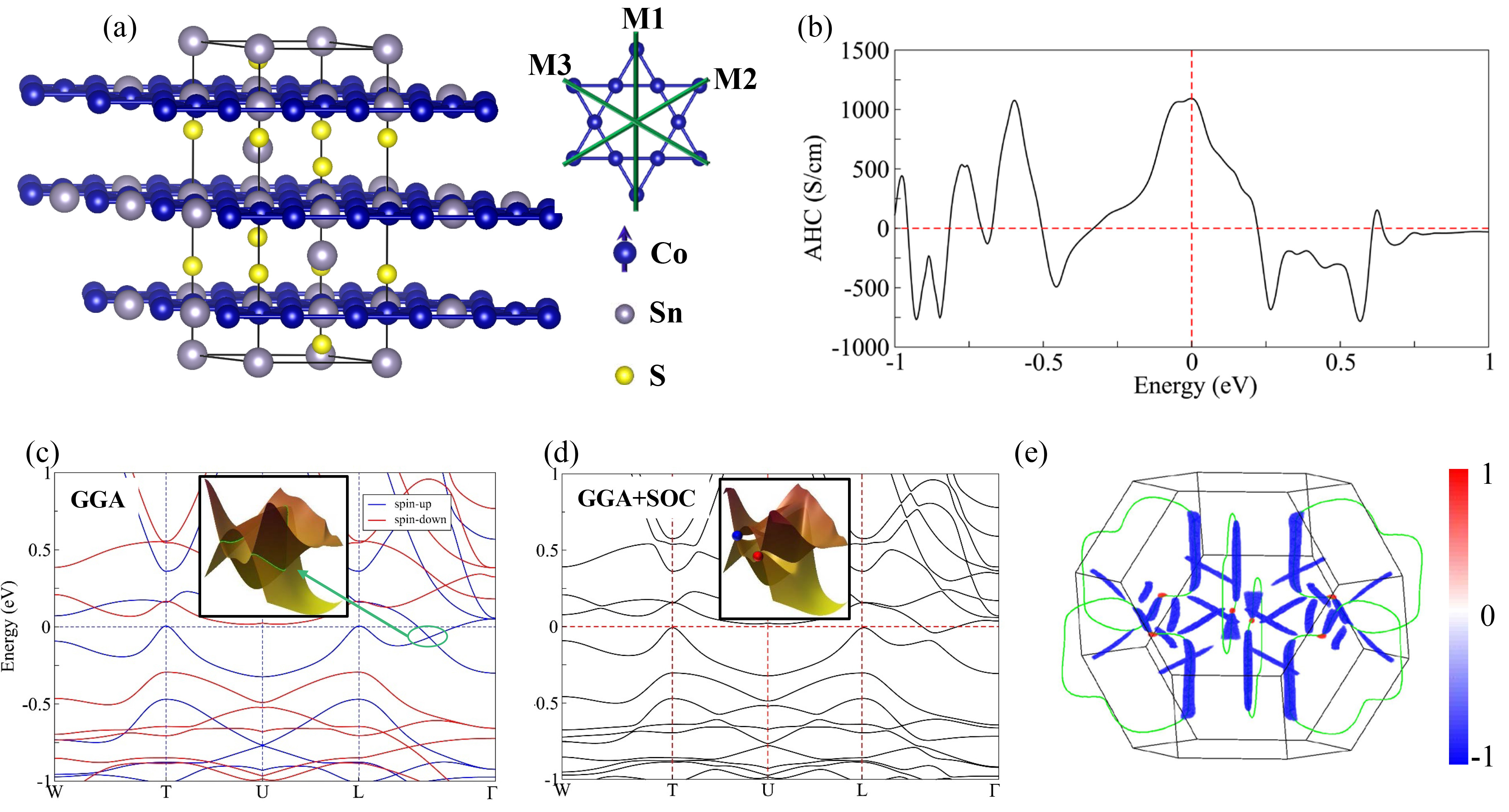}
\caption{
        AHE in the magnetic WSM Co$_3$Sn$_2$S$_2$.
        (\textbf{a}) Crystal and magnetic structure for Co$_3$Sn$_2$S$_2$.
        The~three mirror planes are represented by the green lines. 
        (\textbf{b}) Energy dependent AHC.
        (\textbf{c},\textbf{d}) Energy dispersion along high symmetry lines for the 
        cases without and with SOC, respectively. The~local energy dispersion for the
        nodal line and the Weyl points is shown in the inset. The~green line represents 
        the nodal line. The~blue and red points represent Weyl points.
        (\textbf{e}) Berry curvature distribution in the 3D BZ. 
        The~color bar is in arbitrary units.
}
\label{CSS}
\end{figure}

Taking Co$_3$Sn$_2$S$_2$ as an example, the Co atoms form
 Kagome lattices stacked along the $c$ direction (see Figure~\ref{CSS}a).
Here, the crucial symmetries are the three mirror planes parallel to the $c$ direction,
which results in three pairs of nodal lines connected by a $C_3^z$ rotation
symmetry, as presented in Figure~\ref{CSS}c,e. Because the magnetization is aligned
along the $z$ direction, the mirror symmetries are broken by SOC, and the nodal lines are not longer protected. Therefore, a band gap opens. Meanwhile,
one pair of Weyl points with opposite chirality remains
along each of the former nodal lines  (see Figure~\ref{CSS}d). The~Berry curvature 
distribution in $k$ space is dominated by the SOC-gapped nodal line  
(see Figure~\ref{CSS}e), leading to an AHC of $\sim$1100 S/cm. Together with the
low charge carrier density, the AHA can reach up to $\sim$20\%, which is a
new record in 3D materials~\cite{wang2018,liu2018}.
\section{Anomalous Hall Effect in Antiferromagnetic Topological Semimetals}

From the symmetry point of view, the AHE in 
ferromagnets is easy to understand due to the net
magnetization, and the important insight is that the
special topological band structure can enhance the signal
dramatically. However, for a long time, it was believed that the AHE
cannot exist in antiferromagnets (AFMs), because the measured
Hall resistivity normally follows
$\rho_{Hall}=R_{0}\mu_{0}H+(\alpha\rho_{xx}^{2}+\beta\rho_{xx})$,
where the fist term is the ordinary Hall from the Lorentz
force and the second term is the anomalous component
from extrinsic scattering and intrinsic contributions.

Because the intrinsic contribution can be understood
from the integral of the Berry curvature in $k$ space,
and the Berry curvature is odd with respect to time reversal
symmetry, the AHE can only exist in systems with broken time reversal symmetry. In most collinear AFMs, although the time
reversal symmetry $\hat{T}$ is broken, it is possible to find a combined symmetry
of $\hat{T}$ and a space group operation $\hat{O}$,  which is preserved in the system and still
reverses the sign of Berry curvature. Therefore, the AHE is forced
to zero in these systems. However, such joint symmetry $\hat{T}\hat{O}$ does not
necessarily exist in all AFM systems. Two counter-examples are
non-collinear AFMs and compensated ferrimagnets.

The existence of an AHE in AFMs was first proposed by R. Shindou and 
N. Nagaosa in 2001 by predicting a non-zero AHE in non-coplanar
AFM face-centered-cubic lattices via Berry curvature 
analysis~\cite{shindou2001}. 
Further, it was predicted in 2014 via Berry curvature calculations in cubic Mn$_3$Ir~\cite{chen2014} and hexagonal Mn$_3X$ ($X$ = Ga, Sn or Ge)~\cite{kubler2014} that the intrinsic AHE can even exist in coplanar 
non-collinear AFMs. The~latter was experimentally observed in Hall measurements
soon after the theoretical prediction~\cite{nakatsuji2015,nayak2016}.

\begin{figure}[h]
\centering
\includegraphics[width=0.90\textwidth]{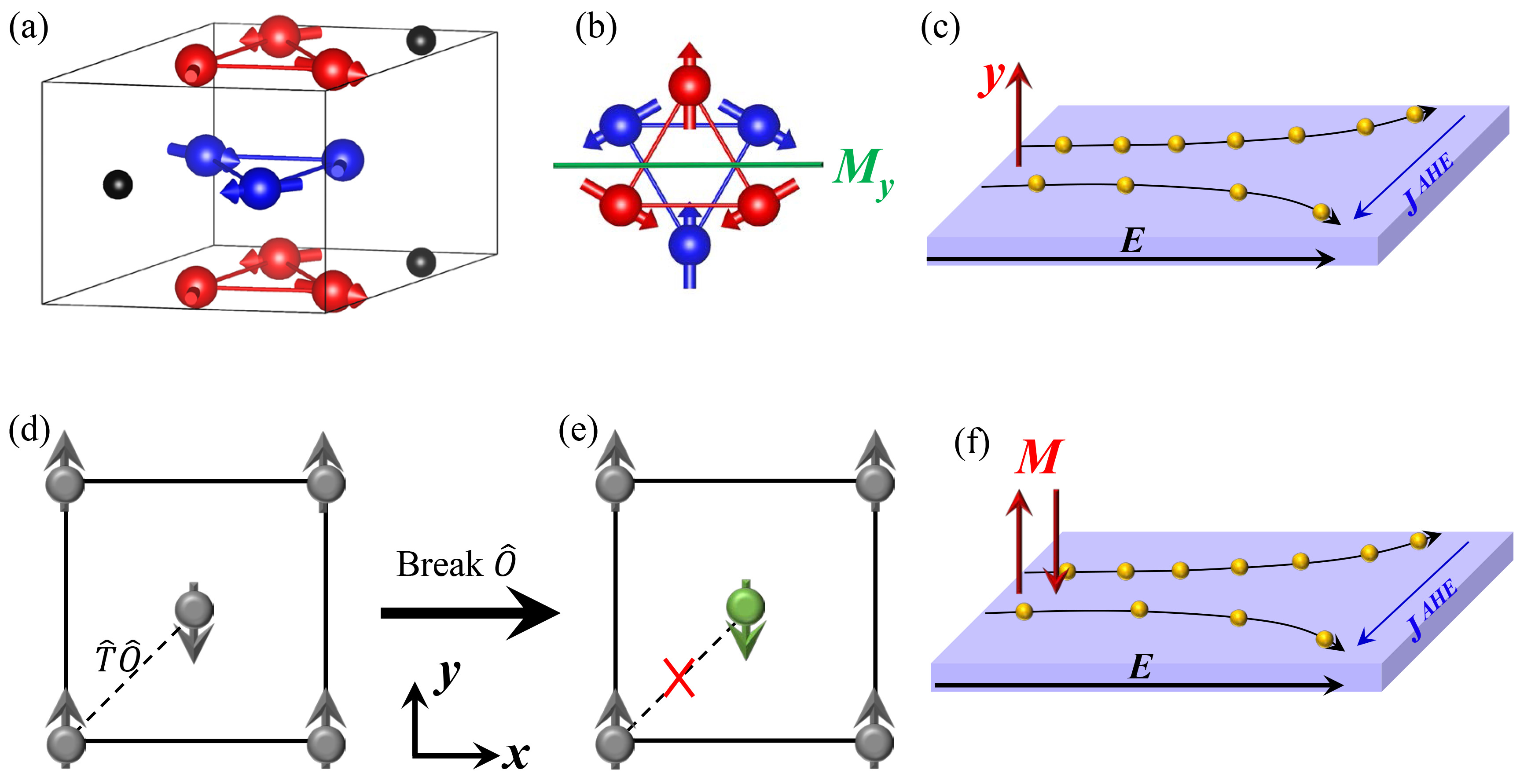}
\caption{
        AHE in AFMs.
        (\textbf{a}) Crystal and magnetic structure of Mn$_3$Ge. The~magnetic moments
        are on the Mn sites. Each Mn layer forms a triangular lattice.
        (\textbf{b}) The~two Mn layers are connected by a glide mirror symmetry.
        The~mirror plane is represented by the green line.
        (\textbf{c}) Schematic of the AHE in Mn$_3$Ge and Mn$_3$Sn. The~$y$-component
        of the AHC is non-zero.
        (\textbf{d}) Collinear AFM with combined time reversal and translational symmetry.
        (\textbf{e}) The~translational symmetry between the two sub-lattices is broken by atomic 
        replacement.
        (\textbf{f}) Schematic of the AHE in compensated ferrimagnets. Only the component along the
        magnetization axis exists.
}
\label{Mn3Ge}
\end{figure}

Recently, it was found that the AHE in Mn$_3$Ge and Mn$_3$Sn
is strongly dependent on the symmetry of the magnetic structure and the
topological band structure~\cite{zhang2017}. 
As presented in Figure~\ref{Mn3Ge}a, the triangular lattices 
formed by Mn are stacked along the $c$ direction,
and the magnetic structure follows a glide mirror symmetry 
$\left\{M_y|(0,0,1/2) \right\}$, see Figure~\ref{Mn3Ge}b. 
Because the Berry curvature is 
invariant with respect to glide operations, the shape of the AHC tensor
is only 
decided by the mirror $M_y$. $M_y$ is similar to a time reversal
operation for the $x$ and $z$ components of the spin and 
will change the sign of $\Omega^{x}$ and $\Omega^{z}$ while 
keeping $\Omega^{y}$ invariant. As a result, the AHC vector 
takes the form $\overrightarrow{\sigma}^{AHE}=\{\begin{array}{ccc}
0, & \sigma^{y}, & 0\end{array}\}$, with only one non-zero
component  (see Figure~\ref{Mn3Ge}c). In addition, it was found that
the size of $\sigma^{y}$ is related to the Weyl points in
Mn$_3$Ge and Mn$_3$Sn. 
The strong AHE in Mn$_3$Ge and Mn$_3$Sn inspired the investigation of their band structures from a
topological point of view.
It was found that there are six pairs of Weyl points 
near the Fermi level in Mn$_3$Sn according to first 
principles calculations~\cite{yang2017},
which were experimentally observed in ARPES measurements soon
after the theoretical prediction~\cite{kuroda2017evidence}.

Apart from a non-collinear magnetic structure, the combined symmetry
that reverses the sign of the Berry curvature can also be broken in
collinear magnetic structure with zero net moment. As~presented 
in Figure~\ref{Mn3Ge}d, the sign of the magnetic moments is changed by the combined
time reversal and translation symmetry. This translation symmetry
can be broken by an atomic replacement while keeping local moments
invariant (see Figure~\ref{Mn3Ge}e). In this case, the system can obtain a
non-zero net Berry curvature and therefore a finite intrinsic 
AHC. This kind of magnetic structure can exist in compensated 
ferrimagnets. Because the charge carrier density is 
normally very small in compensated ferrimagnets, the signal
of the AHE is very weak. However, if there are some special 
topological band structures, such as nodal lines and Weyl points,
the AHE should be strongly enhanced. A possible candidate is
the Heusler compound Ti$_2$MnAl~\cite{shi2018prediction}, 
where the zero net magnetic
moment is due to the opposite magnetization from Ti and Mn sites.

{\section{Anomalous Hall Effect in Thin Films}}
{In the above sections, we only focus  on the AHE in bulk crystals. However, in microelectronic applications, thin films are used most of the time. When discussing thin films from a theory point of view, one has to distinguish between two cases: On the one hand, the film can be above a certain thickness threshold and behave as the bulk system. On the other hand, the film can be of very small thickness. In this case, the bulk picture is not longer valid and extrinsic effects are more pronounced. The~thickness threshold is a material dependent  value, e.g., $\approx$80 nm in Co$_2$MnGa~\cite{PhysRevB.100.054422}.}

\textcolor{black}{When the film thickness is below the threshold value, surface effects get stronger compared to the bulk properties. This can lead to a larger extrinsic contribution to the AHE due to enhanced scattering at the surfaces. Depending on the sign of the extrinsic part with respect to the intrinsic one, this can either enhance or decrease the intrinsic AHE~\cite{changxu2016,rajendra,PhysRevB.100.054445}. For very thin films, it can even be possible that the confinement to leads a changed band structure where the Weyl points are annihilated and an inverted band gap is created~\cite{PhysRevLett.107.186806,PhysRevB.96.045307,muechler2017realization}, leading to a 2D QAHE system.}

\section{Anomalous Nernst Effect in Topological Semimetals}

\begin{figure}[h]
\begin{center}
\includegraphics[width=0.90\textwidth]{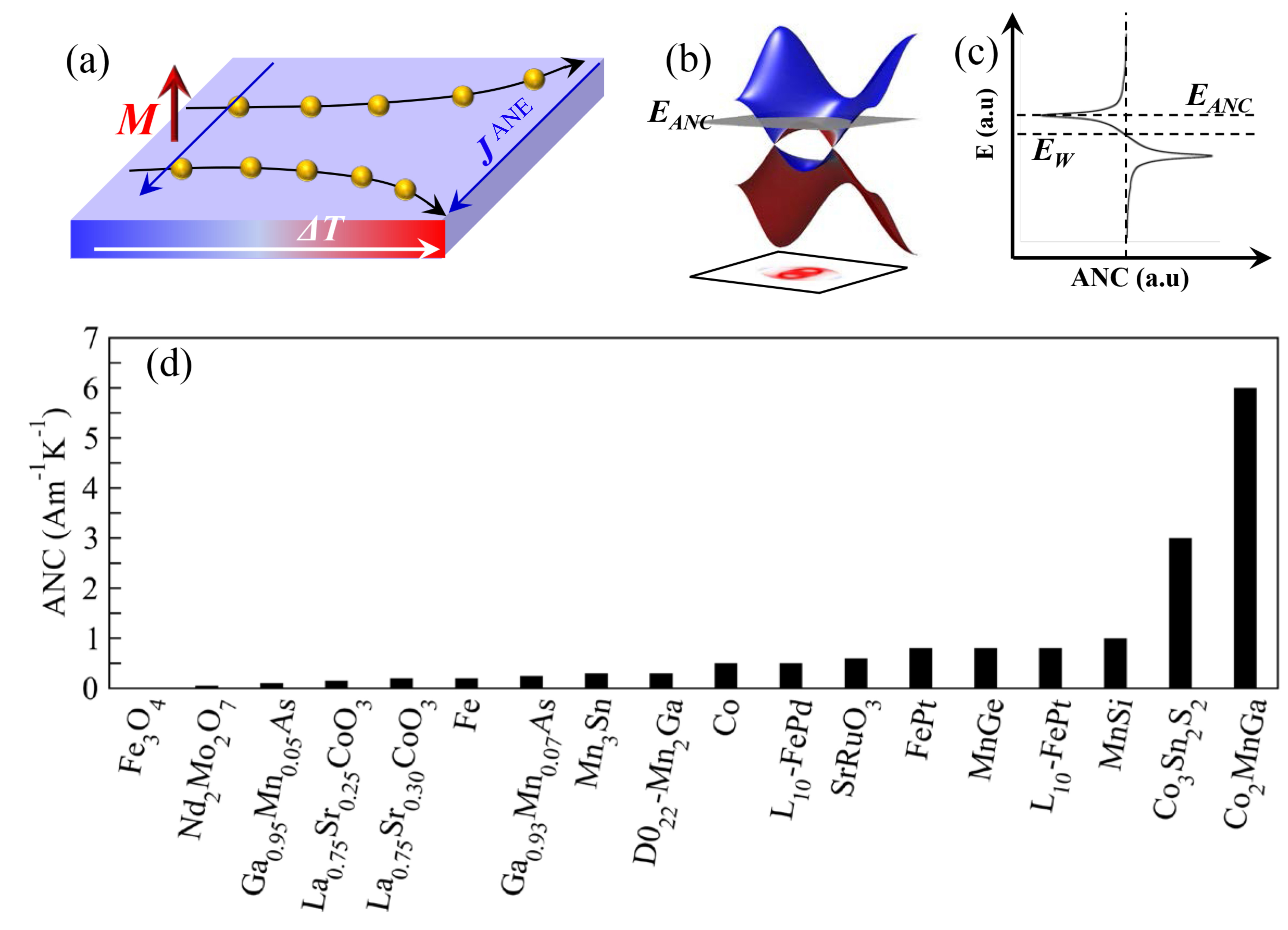}
\end{center}
\caption{
        ANE in magnetic materials.
        (\textbf{a}) Schematic of the ANE.
        (\textbf{b}) Energy dispersion of an effective model for a magnetic WSM.
        The~gray plane represents the energy with the peak value of the ANC.
        (\textbf{c}) Energy dependent ANC.
        (\textbf{d}) Absolute value of the ANC for different magnetic metals, with data
	collected from Refs.~\cite{Hirokane2016,Miyasato2007,Ikhlas2017,Li2017,Hanasaki2008,Pu2008,Ramos2014,Weischenberg2013}
.}
\label{ane}
\end{figure}

Apart from the AHE, there is also the corresponding thermoelectric effect, the
anomalous Nernst effect (ANE), which can be utilized for measurements
of Berry phase effects in band structures. In the ANE, a
transverse electrical current in magnetic materials is generated by an applied longitudinal temperature
gradient, instead of an electric field~\cite{bauer2012spin,lee2004anomalous,xiao2006berry}, 
as presented in Figure~\ref{ane}a.
Because also the ANE is dependent on the Berry curvature,
a strong ANE is expected in magnetic topological materials.
Very recently, large ANEs were observed in the
magnetic WSMs Co$_2$MnGa~\cite{sakai2018giant,guin2019anomalous} 
and Co$_3$Sn$_2$S$_2$~\cite{guin2019zero}, which are believed to
originate from the Weyl points and the nodal line band structures.
Owing to the contribution of the topological band structure, the
anomalous Nernst conductivity (ANC) in Co$_2$MnGa and Co$_3$Sn$_2$S$_2$
can reach values one order of magnitude larger than in
traditional magnetic materials (see Figure~\ref{ane}d).

Although AHE and ANE share the same symmetry and both    
can be understood from the Berry curvature, their dependence
on the Berry curvature is different~\cite{xiao2006berry}. 
The ANC is generated by
the combination of Berry curvature ($\Omega$), distribution function
$f(n)$, and temperature $T$ in the form of 
\begin{equation}
\begin{aligned}
\alpha^{i}=\frac{1}{T}\frac{e}{\hbar}\sum_{k}\sum_{n}\Omega^{i}[(E_{n}-E_{F})f_{n}+k_{B}T\ln\left(1+exp\frac{E_{n}-E_{F}}{-k_{B}T}\right)].
\end{aligned}
\label{ANEeq}
\end{equation}

Similar to other thermoelectric signals, the ANE is more sensitive to 
the first derivative of the electrical characteristics with respect 
to energy, rather than to the value itself. As discussed above, 
the AHC shows a peak value at the energy of the Weyl points in 
high symmetric WSMs due to the big contribution
from the Weyl points. In contrast to that, the ANC is zero
at the energy of the Weyl points  (see Figure~\ref{ane}b,c), and two peak values
appear away from that energy. Owing to the different mechanism
in the ANE, it can be used to detect topological band structure features
away from the Fermi level by the means of thermoelectric
transport measurements, even if the electronic transport signal 
is weak~\cite{noky2018characterization}.

\section{Spin Hall Effect in Topological Materials}

\begin{figure}[h]
\centering
\includegraphics[width=0.90\textwidth]{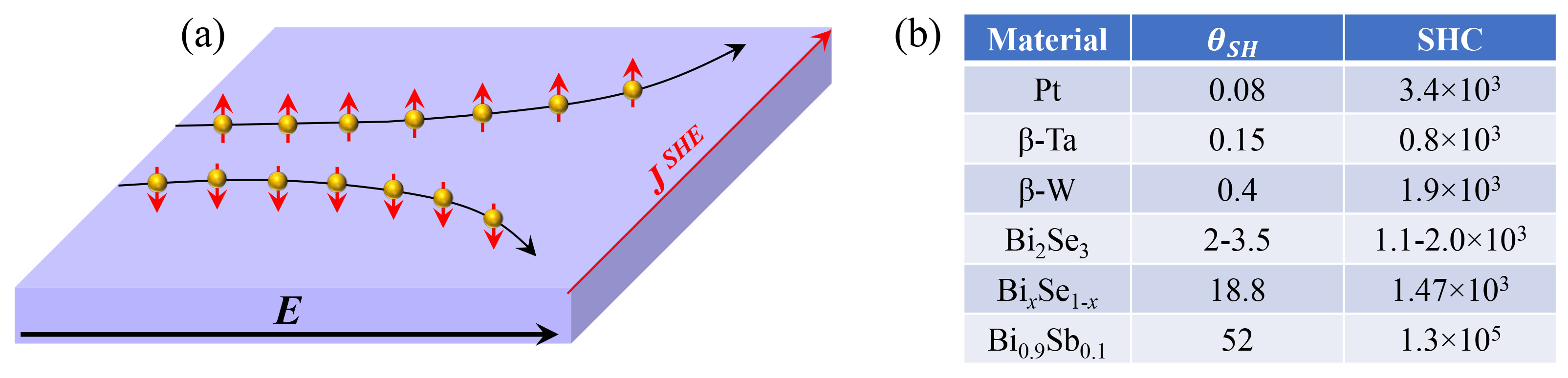}
\caption{
        SHE in topological materials.
        (\textbf{a}) Schematic of the SHE. A transverse pure spin current is generated
        by an applied longitudinal electric field.
        (\textbf{b}) The~table for some typical SHE materials, with data collected 
	from Refs.~\cite{Liu2012,Pai2012,Liu2011,mellnik2014spin,mahendra2018room,khang2018conductive}.
	The unit of SHC is $\frac{e}{2\hbar}\frac{S}{cm}$.
}
\label{SHE}
\end{figure}

Because the Berry curvature is odd with respect to the time reversal 
operation, the AHE can only exist in magnetic systems. 
However, although the AHE is forbidden, the spin Hall effect 
(SHE) is allowed in time reversal symmetric systems. In the SHE, a transversal spin current is generated by 
an applied longitudinal spin 
current due to the SOC~\cite{d1971possibility,hirsch1999spin,kato2004observation,sinova2015spin}  
(see Figure~\ref{SHE})a. Conversely, a transverse
charge current can be generated via a longitudinal spin
current. Hence, the SHE provides an effective method to
generate and manipulate the spin current without a magnetic
field.

The extrinsic SHE was proposed by M. I. Dyakonov and 
V. I. Perel in 1971~\cite{d1971possibility}, 
and the SHE received extensive attention after the
theoretical studies of its intrinsic 
mechanism~\cite{Murakami2003,sinova2004universal,Bernevig2005} 
and its experimental observations~\cite{kato2004observation,Wunderlich2005,Day2005}.
The study of the quantum version of the SHE led to the
2D TI~\cite{Kane2005,Kane2005b,bernevig2006quantum}, 
and the spin current from
spin--momentum locked topological surface states  
opens up new approaches to the SHE.

Similar to the AHC, the spin Hall conductivity (SHC)
can be formulated via the spin Berry curvature
in the Kubo formula approach~\cite{Murakami2003,sinova2004universal,Bernevig2005}, 
where the observable is the spin current density
with $\hat{A}=\frac{-e}{2\hbar}\frac{\{\hat{\overrightarrow{s}},\hat{\overrightarrow{v}}\}}{V}$.
The spin current $\overrightarrow{J}_{s-i}^{k}$
is generated by the applied electric field
$\overrightarrow{E}_{j}$ following the linear
response
$\overrightarrow{J}_{s-i}^{k}=\sigma_{ij}^{k}\overrightarrow{E}_{j}$. Here, the SHC tensor $\sigma_{ij}^{k}$ describes a spin current along $i$ with
spin polarization along $k$ and an applied
field in the $j$ direction. Because the AHC and SHC share a
similar formalism, the linear band crossings and
anti-crossings can also generate strong local spin
Berry curvature in both magnetic and nonmagnetic
systems. Additionally, because the SHE originates
from the SOC, a strong intrinsic SHE is expected
in both TI and topological metals.

In SHE experiments, apart from the SHC, there is another important
parameter, the spin Hall angle (SHA). It represents the efficiency 
of conversion from charge current to spin current.
Thus far, there are mainly two approaches for SHE materials:
one is based on heavy transition 
metals~\cite{Tanaka2008, Hoffmann2013,sinova2015spin, Kimura2007, Saitoh2006}, 
and the other is based on TIs or topological 
semimetals~\cite{mellnik2014spin,khang2018conductive,fan2014magnetization,mahendra2018room}. The~SHC and SHA for 
some reported SHE materials are given in Figure~\ref{SHE}b. Very 
recently, it was found that the nodal line band structures play 
a crucial role for a strong SHE even in transition metals~\cite{zhangyang2019}.

\section{Spin Nernst Effect in Topological Materials}

The relationship between AHE and ANE can also be transferred to spin current generation. In~addition to SHE where the driving 
force is an electric field, a spin current can   also  be
generated by a temperature gradient, which is known as the spin Nernst 
effect (SNE)~\cite{Meyer2017,sheng2017spin,kim2017observation}. 
In the SNE, a transverse
spin current $\overrightarrow{J}_{s-i}^{k}$ can be generated by an applied temperature
gradient $\nabla\overrightarrow{T}_{j}$ via the linear response
$\overrightarrow{J}_{s-i}^{k}=\alpha_{ij}^{k}\nabla\overrightarrow{T}_{j}$,
where $\alpha_{ij}^{k}$ is the linear response tensor of the spin Nernst
conductivity (SNC). Very recently, the SNE was observed in heavy transition
metals in thermoelectric transport measurements~\cite{Meyer2017,sheng2017spin,kim2017observation}.

\begin{figure}[h]
\centering
\includegraphics[width=0.90\textwidth]{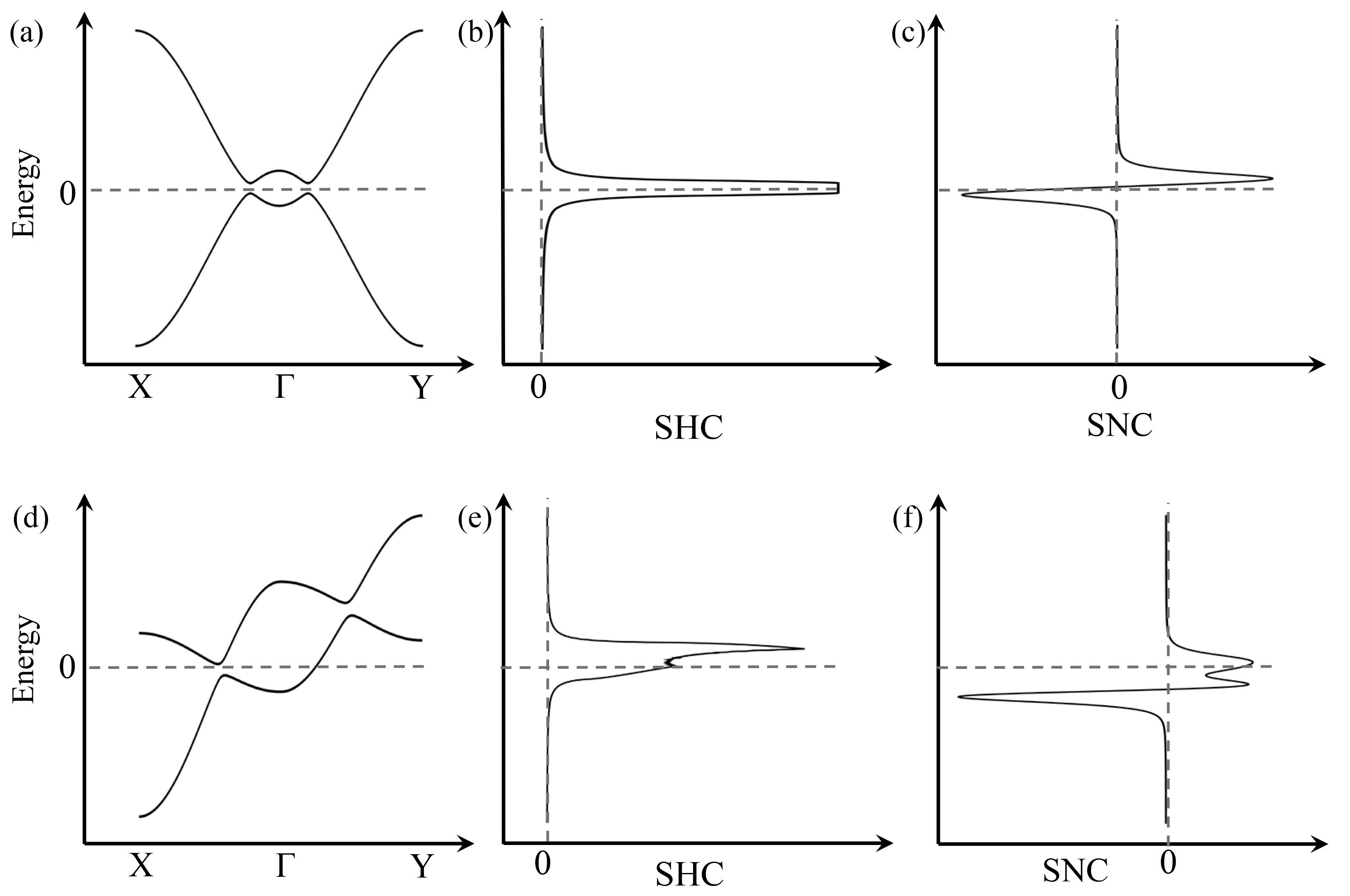}
\caption{
SNE in TIs and topological semimetals.
        (\textbf{a}) Energy dispersion of a TI.
        (\textbf{b},\textbf{c}) Energy dependent SHC and SNC for a TI, respectively.
        (\textbf{d}) Energy dispersion for a topological semimetal with tilted
        band structure.
        (\textbf{e},\textbf{f}) Energy dependent SHC and SNC for a topological semimetal, respectively.
}
\label{SNE}
\end{figure}

Because the SNE shares similar mechanisms with the ANE, only  with 
different symmetry requirements, a strong SNE is also 
expected in topological materials. Figure~\ref{SNE} gives
a general understanding of the relation between 
SNE and topological band structures. For a symmetry 
model of a TI with a global band gap, a constant
SHC exists in the inverted band gap. However, 
because the SNE is dependent on the first derivative of the 
SHE with respect to the chemical potential, the SNE vanishes
at the charge neutral point, as presented in Figure~\ref{SNE}a--c. 
To obtain a non-zero finite SNE, breaking the 
balance of the SHC distribution in energy space is necessary. 
An effective way to achieve this is to decrease the symmetry and
tilt the inverted band gap with a finite Fermi surface, 
as the schematically shown
in Figure~\ref{SNE}d. With the tilted inverted band gap, the distribution of SHC 
is no longer  symmetric   in energy space and therefore a finite SNC appears at the 
Fermi energy  (see Figure~\ref{SNE}e,f). From the model analysis, one can find that
a topological semimetal or a doped topological insulator is expected to host a
strong SNE, which is similar to the ANE in magnetic topological semimetals. 

\section{Spin--Orbit Torque in Magnetic Semimetals}

Further breaking inversion symmetry in magnetic materials, 
one can have another linear response effect  
to control the magnetization orientation by electrical 
current~\cite{manchon2009theory,garate2009influence}. 
Utilizing the inverse spin--galvanic effect in a 
non-centrosymmetric magnetic 
system, the non-equilibrium spin polarization will interact via exchange-coupling with the magnetization and acts as an 
effective field. This effective field originates 
from the spin--orbit coupling and can 
generate torques on the magnetic moments, which allows the control of the magnetic states by an electric field. Therefore, it is 
called spin--orbit torque~(SOT). 

Most SOT studies have focused on ferromagnetic 
systems where the inversion symmetry is broken by either the 
bulk crystal unit 
cell~\cite{chernyshov2009evidence,kurebayashi2014antidamping} 
or due to heterostructures~\cite{miron2010current,miron2011perpendicular,liu2012current, Pi2010, Liu2012,KimJunyoen2013,Suzuki2011}. As AFMs
are much less sensitive to external magnetic fields
and their dynamics of the magnetic moments are much faster 
than those in FMs, spintronics in AFMs have attracted 
intensive attention~\cite{Gomonay2014,MacDonald2011,Jungwirth2016}. 
Because the manipulation of 
the magnetic order through the traditional approach 
of using an external field requires too large fields, 
the electrical manipulation of AFMs via SOT is much 
more desirable. Recently, a SOT phenomenon has been 
theoretically proposed~\cite{Zelezny2014} and experimentally observed 
in bulk collinear antiferromagnetic Dirac semimetal
CuMnAs~\cite{Wadley2016,Tang2016}. 
In CuMnAs, the inversion symmetry is broken by magnetic 
order, but the band spin degeneracy is protected by joint 
inversion and time-reversal symmetry. In such an AFM, 
an electric current can generate a staggered torque 
for each sub-lattice, which is an effective way to manipulate
the magnetic order~\cite{Smejkal2017}.

\begin{figure}[h]
\centering
\includegraphics[width=0.90\textwidth]{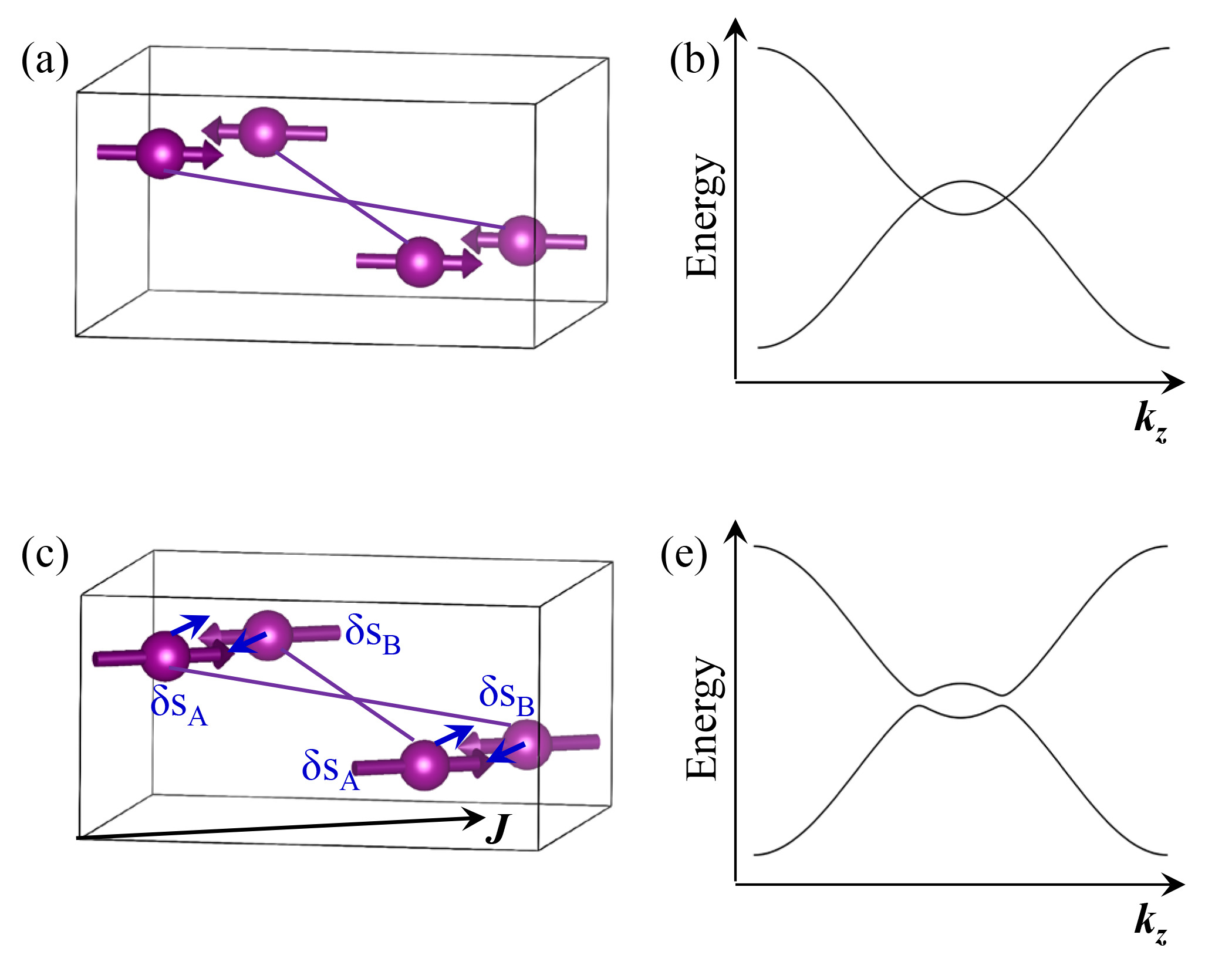}
\caption{
        SOT in AFM CuMnAs.
        (\textbf{a}) Magnetic structure for orthorhombic CuMnAs with
        spin polarization along the $z$-axis.
        (\textbf{b}) Energy dispersion of CuMnAs. It is a Dirac
        semimetal when spin polarization is along the $z$-axis.
        (\textbf{c}) The~spin orientation is moved away from the $z$-axis
        by an applied electric field via the SOT. The~torques
        for the two spin sub-lattice A and B are opposite.
        (\textbf{d})~The~Dirac point gets a mass term and a band gap opens
        when the spin polarization moves away from the $z$~direction.
}
\label{sot}
\end{figure}

CuMnAs can exist in  
both orthorhombic and tetragonal crystal structures in experiments~\cite{Wadley2016,Tang2016,Smejkal2017},
and both     can host the antiferromagnetic Dirac semimetal phase.
Taking the orthorhombic phase as an example, four Mn sites form
two spin sub-lattices, which are connected by the combined $\hat{P}\hat{T}$
symmetry, as presented in Figure~\ref{sot}a,c. When the spin polarization is
along $[001]$ direction, the crystal has a screw rotation 
symmetry $S_{z}=\{C_{2z}|(0.5,0,0.5)\}$,
in which the bands with opposite $S_z$ eigenvalues guarantee the 
linear band crossing in the inverted band structure  (see Figure~\ref{sot}b). 
The spin orientation can be tuned by an
applied electrical current via the SOT, as presented in Figure~\ref{sot}c. As the
spin polarization moves away from $[001]$-axis, the screw rotation symmetry
is broken, and the Dirac point obtains a non-zero mass  (see Figure~\ref{sot}d). 
Therefore, the SOT provides an effective method to tune the topological 
band structures in magnetic topological states.

\section{Summary}
 We introduce  the linear response effects in topological 
materials based on fundamental theories and recent developments of 
experiments. The~interplay of symmetry and electromagnetic structures
tells us how to obtain corresponding materials with non-zero 
linear response signals. The~topological 
band theory offers a direction to achieve enhanced and even quantum
linear response effects, which is essential to find the correct
materials for the next generation of technological~applications.

\section{Author contributions}
Conceptualization, S.Y. and J.N.; writing, J.N.; and supervision, S.Y.

\section{Conflicts of interest}
The authors declare no conflict of interest.


\begin{thebibliography}{999}

\bibitem[Kane and Mele(2005{\natexlab{a}})]{Kane2005}
Kane, C.L.; Mele, E.J.
\newblock {Z2 Topological Order and the Quantum Spin Hall Effect}.
\newblock {\em Phys. Rev. Lett.} {\bf 2005}, {\em 95},~146802.

\bibitem[Kane and Mele(2005{\natexlab{b}})]{Kane2005b}
Kane, C.L.; Mele, E.J.
\newblock Quantum Spin Hall Effect in Graphene.
\newblock {\em Phy. Rev. Lett.} {\bf 2005}, {\em 95},~226801.

\bibitem[Bernevig \em{et~al.}(2006)Bernevig, Hughes, and
  Zhang]{bernevig2006quantum}
Bernevig, B.A.; Hughes, T.L.; Zhang, S.C.
\newblock {Quantum spin Hall effect and topological phase transition in HgTe
  quantum wells}.
\newblock {\em Science} {\bf 2006}, {\em 314},~1757--1761.

\bibitem[K{\"o}nig \em{et~al.}(2007)K{\"o}nig, Wiedmann, Br{\"u}ne, Roth,
  Buhmann, Molenkamp, Qi, and Zhang]{konig2007quantum}
K{\"o}nig, M.; Wiedmann, S.; Br{\"u}ne, C.; Roth, A.; Buhmann, H.; Molenkamp,
  L.W.; Qi, X.L.; Zhang, S.C.
\newblock {Quantum spin Hall insulator state in HgTe quantum wells}.
\newblock {\em Science} {\bf 2007}, {\em 318},~766--770.

\bibitem[Fu \em{et~al.}(2007)Fu, Kane, and Mele]{fu2007topological}
Fu, L.; Kane, C.L.; Mele, E.J.
\newblock {Topological insulators in three dimensions}.
\newblock {\em Phys. Rev. Lett.} {\bf 2007}, {\em 98},~106803.

\bibitem[Hsieh \em{et~al.}(2008)Hsieh, Qian, Wray, Xia, Hor, Cava, and
  Hasan]{hsieh2008topological}
Hsieh, D.; Qian, D.; Wray, L.; Xia, Y.; Hor, Y.S.; Cava, R.J.; Hasan, M.Z.
\newblock {A topological Dirac insulator in a quantum spin Hall phase}.
\newblock {\em Nature} {\bf 2008}, {\em 452},~970.

\bibitem[Hasan and Kane(2010)]{Hasan2010ku}
Hasan, M.Z.; Kane, C.L.
\newblock {Colloquium: Topological insulators}.
\newblock {\em Rev. Mod. Phys.} {\bf 2010}, {\em 82},~3045--3067.

\bibitem[Qi and Zhang(2011)]{Qi2011RMP}
Qi, X.L.; Zhang, S.C.
\newblock {Topological insulators and superconductors}.
\newblock {\em Rev. Mod. Phys.} {\bf 2011}, {\em 83},~1057.

\bibitem[Chiu \em{et~al.}(2016)Chiu, Teo, Schnyder, and
  Ryu]{chiu2016classification}
Chiu, C.K.; Teo, J.C.; Schnyder, A.P.; Ryu, S.
\newblock {Classification of topological quantum matter with symmetries}.
\newblock {\em Rev. Mod. Phys.} {\bf 2016}, {\em 88},~035005.

\bibitem[Po \em{et~al.}(2017)Po, Vishwanath, and Watanabe]{po2017symmetry}
Po, H.C.; Vishwanath, A.; Watanabe, H.
\newblock {Symmetry-based indicators of band topology in the 230 space groups}.
\newblock {\em Nat. Commun.} {\bf 2017}, {\em 8},~50.

\bibitem[Song \em{et~al.}(2018{\natexlab{a}})Song, Zhang, Fang, and
  Fang]{song2018quantitative}
Song, Z.; Zhang, T.; Fang, Z.; Fang, C.
\newblock {Quantitative mappings between symmetry and topology in solids}.
\newblock {\em Nat. Commun.} {\bf 2018}, {\em 9},~3530.

\bibitem[Song \em{et~al.}(2018{\natexlab{b}})Song, Zhang, and
  Fang]{song2017diagnosis}
Song, Z.; Zhang, T.; Fang, C.
\newblock {Diagnosis for Nonmagnetic Topological Semimetals in the Absence of
  Spin-Orbital Coupling}.
\newblock {\em Phys. Rev. X} {\bf 2018}, {\em 8},~031069.
\newblock
  doi:{\href{https://doi.org/10.1103/PhysRevX.8.031069}{\detokenize{10.1103/PhysRevX.8.031069}}}.

\bibitem[Kitaev(2009)]{kitaev2009periodic}
Kitaev, A.
\newblock {Periodic table for topological insulators and superconductors}.
\newblock  \emph{AIP Conf. Proc.}   \textbf{2009},  \emph{1134},  22--30.

\bibitem[Fu(2011)]{fu2011topological}
Fu, L.
\newblock {Topological crystalline insulators}.
\newblock {\em Phys. Rev. Lett.} {\bf 2011}, {\em 106},~106802.

\bibitem[Wan \em{et~al.}(2011)Wan, Turner, Vishwanath, and Savrasov]{Wan2011}
Wan, X.G.; Turner, A.M.; Vishwanath, A.; Savrasov, S.Y.
\newblock {Topological semimetal and Fermi-arc surface states in the electronic
  structure of pyrochlore iridates}.
\newblock {\em Phys. Rev. B} {\bf 2011}, {\em 83},~205101.

\bibitem[Burkov and Balents(2011)]{Burkov2011de}
Burkov, A.A.; Balents, L.
\newblock {Weyl Semimetal in a Topological Insulator Multilayer}.
\newblock {\em Phys. Rev. Lett.} {\bf 2011}, {\em 107},~127205.

\bibitem[Wang \em{et~al.}(2012)Wang, Sun, Chen, Franchini, Xu, {Hongming Weng},
  Dai, and {Zhong Fang}]{Wang2012}
Wang, Z.; Sun, Y.; Chen, X.Q.; Franchini, C.; Xu, G.; {Hongming Weng}.; Dai,
  X.; {Zhong Fang}.
\newblock {Dirac semimetal and topological phase transitions in A$_3$Bi (A= Na,
  K, Rb)}.
\newblock {\em Phys. Rev. B} {\bf 2012}, {\em 85},~195320.

\bibitem[Hsieh \em{et~al.}(2012)Hsieh, Lin, Liu, Duan, Bansil, and
  Fu]{hsieh2012topological}
Hsieh, T.H.; Lin, H.; Liu, J.; Duan, W.; Bansil, A.; Fu, L.
\newblock {Topological crystalline insulators in the SnTe material class}.
\newblock {\em Nat. Commun.} {\bf 2012}, {\em 3},~982.

\bibitem[Young \em{et~al.}(2012)Young, Zaheer, Teo, Kane, Mele, and
  Rappe]{Young2012}
Young, S.M.; Zaheer, S.; Teo, J.C.Y.; Kane, C.L.; Mele, E.J.; Rappe, A.M.
\newblock {Dirac Semimetal in Three Dimensions}.
\newblock {\em Phys. Rev. Lett.} {\bf 2012}, {\em 108},~140405.

\bibitem[Ando and Fu(2015)]{ando2015topological}
Ando, Y.; Fu, L.
\newblock {Topological crystalline insulators and topological superconductors:
  from concepts to materials}.
\newblock {\em Annu. Rev. Condens. Matter Phys.} {\bf 2015}, {\em 6},~361--381.

\bibitem[Weng \em{et~al.}(2015)Weng, Fang, Fang, Bernevig, and Dai]{Weng2015}
Weng, H.; Fang, C.; Fang, Z.; Bernevig, B.A.; Dai, X.
\newblock {Weyl Semimetal Phase in Noncentrosymmetric Transition-Metal
  Monophosphides}.
\newblock {\em Phys. Rev. X} {\bf 2015}, {\em 5},~011029.

\bibitem[Huang \em{et~al.}(2015)Huang, Xu, Belopolski, Lee, Chang, Wang,
  Alidoust, Bian, Neupane, Zhang, et~al.]{huang2015weyl}
Huang, S.M.; Xu, S.Y.; Belopolski, I.; Lee, C.C.; Chang, G.; Wang, B.;
  Alidoust, N.; Bian, G.; Neupane, M.; Zhang, C.; et al.
\newblock {A Weyl Fermion semimetal with surface Fermi arcs in the transition
  metal monopnictide TaAs class}.
\newblock {\em Nat. Commun.} {\bf 2015}, {\em 6},~7373.

\bibitem[Xu \em{et~al.}(2015)Xu, Belopolski, Alidoust, Neupane, Bian, Zhang,
  Sankar, Chang, Zhujun, Lee, Shin-Ming, Zheng, Ma, Sanchez, Wang, Bansil,
  Chou, Shibayev, Lin, Jia, and Hasan]{Xu2015TaAs}
Xu, S.Y.; Belopolski, I.; Alidoust, N.; Neupane, M.; Bian, G.; Zhang, C.;
  Sankar, R.; Chang, G.; Zhujun, Y.; Lee,~C.C.; et al.
\newblock {Discovery of a Weyl fermion semimetal and topological Fermi arcs}.
\newblock {\em Science} {\bf 2015}, {\em 349},~613.

\bibitem[Lv \em{et~al.}(2015)Lv, Weng, Fu, Wang, Miao, Ma, Richard, Huang,
  Zhao, Chen, Fang, Dai, Qian, and Ding]{Lv2015TaAs}
Lv, B.Q.; Weng, H.M.; Fu, B.B.; Wang, X.P.; Miao, H.; Ma, J.; Richard, P.;
  Huang, X.C.; Zhao, L.X.; Chen,~G.F.; et al.
\newblock {Experimental Discovery of Weyl Semimetal TaAs}.
\newblock {\em Phys. Rev. X} {\bf 2015}, {\em 5},~031013.

\bibitem[Bradlyn \em{et~al.}(2016)Bradlyn, Cano, Wang, Vergniory, Felser, Cava,
  and Bernevig]{bradlyn2016beyond}
Bradlyn, B.; Cano, J.; Wang, Z.; Vergniory, M.; Felser, C.; Cava, R.; Bernevig,
  B.A.
\newblock {Beyond Dirac and Weyl fermions: Unconventional quasiparticles in
  conventional crystals}.
\newblock {\em Science} {\bf 2016}, {\em 353},~5037.

\bibitem[Belopolski \em{et~al.}(2016)Belopolski, Sanchez, Ishida, Pan, Yu, Xu,
  Chang, Chang, Zheng, Alidoust, et~al.]{belopolski2016discovery}
Belopolski, I.; Sanchez, D.S.; Ishida, Y.; Pan, X.; Yu, P.; Xu, S.Y.; Chang,
  G.; Chang, T.R.; Zheng, H.; Alidoust,~N.; et al.
\newblock Discovery of a new type of topological Weyl fermion semimetal state
  in MoxW1-xTe2.
\newblock {\em Nat. Commun.} {\bf 2016}, {\em 7},~13643.

\bibitem[Chang \em{et~al.}({2016})Chang, Xu, Zheng, Singh, Hsu, Bian, Alidoust,
  Belopolski, Sanchez, Zhang, Lin, and Hasan]{changxu2016}
Chang, G.; Xu, S.Y.; Zheng, H.; Singh, B.; Hsu, C.H.; Bian, G.; Alidoust, N.;
  Belopolski, I.; Sanchez, D.S.; Zhang, S.; et al.  
\newblock {Room-temperature magnetic topological Weyl fermion and nodal line
  semimetal states in half-metallic Heusler Co2TiX (X=Si, Ge, or Sn)}.
\newblock {\em {Sci. Rep.}} {\bf {2016}}, {\em {6}}, 38839.

\bibitem[Chang \em{et~al.}(2016)Chang, Xu, Sanchez, Huang, Lee, Chang, Bian,
  Zheng, Belopolski, Alidoust, Jeng, Bansil, Lin, and Hasan]{Change1600295}
Chang, G.; Xu, S.Y.; Sanchez, D.S.; Huang, S.M.; Lee, C.C.; Chang, T.R.; Bian,
  G.; Zheng, H.; Belopolski,~I.; Alidoust, N.; et al.
\newblock A strongly robust type II Weyl fermion semimetal state in Ta3S2.
\newblock {\em Sci. Adv.} {\bf 2016}, {\em 2},~e1600295.

\bibitem[Bradlyn \em{et~al.}(2017)Bradlyn, Elcoro, Cano, Vergniory, Wang,
  Felser, Aroyo, and Bernevig]{bradlyn2017topological}
Bradlyn, B.; Elcoro, L.; Cano, J.; Vergniory, M.; Wang, Z.; Felser, C.; Aroyo,
  M.; Bernevig, B.A.
\newblock {Topological quantum chemistry}.
\newblock {\em Nature} {\bf 2017}, {\em 547},~298.

\bibitem[Zheng \em{et~al.}(2017)Zheng, Chang, Huang, Guo, Zhang, Zhang, Yin,
  Xu, Belopolski, Alidoust, Sanchez, Bian, Chang, Neupert, Jeng, Jia, Lin, and
  Hasan]{PhysRevLett.119.196403}
Zheng, H.; Chang, G.; Huang, S.M.; Guo, C.; Zhang, X.; Zhang, S.; Yin, J.; Xu,
  S.Y.; Belopolski, I.; Alidoust, N.; et al.
\newblock Mirror Protected Dirac Fermions on a Weyl Semimetal NbP Surface.
\newblock {\em Phys. Rev. Lett.} {\bf 2017}, {\em 119},~196403.
\newblock
  doi:{\href{https://doi.org/10.1103/PhysRevLett.119.196403}{\detokenize{10.1103/PhysRevLett.119.196403}}}.

\bibitem[Fei \em{et~al.}(2017)Fei, Bo, Wang, Wu, Jiang, Fu, Gao, Zheng, Chen,
  Wang, Bu, Song, Wan, Wang, and Wang]{PhysRevB.96.041201}
Fei, F.; Bo, X.; Wang, R.; Wu, B.; Jiang, J.; Fu, D.; Gao, M.; Zheng, H.; Chen,
  Y.; Wang, X.; et al.
\newblock Nontrivial Berry phase and type-II Dirac transport in the layered
  material $\mathrm{PdT}{\mathrm{e}}_{2}$.
\newblock {\em Phys. Rev. B} {\bf 2017}, {\em 96},~041201.
\newblock
  doi:{\href{https://doi.org/10.1103/PhysRevB.96.041201}{\detokenize{10.1103/PhysRevB.96.041201}}}.

\bibitem[Xu \em{et~al.}(2017)Xu, Alidoust, Chang, Lu, Singh, Belopolski,
  Sanchez, Zhang, Bian, Zheng, Husanu, Bian, Huang, Hsu, Chang, Jeng, Bansil,
  Neupert, Strocov, Lin, Jia, and Hasan]{Xue1603266}
Xu, S.Y.; Alidoust, N.; Chang, G.; Lu, H.; Singh, B.; Belopolski, I.; Sanchez,
  D.S.; Zhang, X.; Bian, G.; Zheng, H.; et al.
\newblock Discovery of Lorentz-violating type II Weyl fermions in LaAlGe.
\newblock {\em Sci. Adv.} {\bf 2017}, {\em 3},
\newblock
  doi:{\href{https://doi.org/10.1126/sciadv.1603266}{\detokenize{10.1126/sciadv.1603266}}}.

\bibitem[Armitage \em{et~al.}(2018)Armitage, Mele, and
  Vishwanath]{armitage2018weyl}
Armitage, N.; Mele, E.; Vishwanath, A.
\newblock {Weyl and Dirac semimetals in three-dimensional solids}.
\newblock {\em \mbox{Rev. Mod. Phys.}} {\bf 2018}, {\em 90},~015001.

\bibitem[Schindler \em{et~al.}(2018)Schindler, Cook, Vergniory, Wang, Parkin,
  Bernevig, and Neupert]{schindler2018higher}
Schindler, F.; Cook, A.M.; Vergniory, M.G.; Wang, Z.; Parkin, S.S.; Bernevig,
  B.A.; Neupert, T.
\newblock {Higher-order topological insulators}.
\newblock {\em Sci. Adv.} {\bf 2018}, {\em 4},~0346.

\bibitem[Zhang \em{et~al.}(2019)Zhang, Jiang, Song, Huang, He, Fang, Weng, and
  Fang]{zhang2019catalogue}
Zhang, T.; Jiang, Y.; Song, Z.; Huang, H.; He, Y.; Fang, Z.; Weng, H.; Fang, C.
\newblock {Catalogue of topological electronic materials}.
\newblock {\em Nature} {\bf 2019}, {\em 566},~475.

\bibitem[Tang \em{et~al.}(2019)Tang, Po, Vishwanath, and
  Wan]{tang2019comprehensive}
Tang, F.; Po, H.C.; Vishwanath, A.; Wan, X.
\newblock {Comprehensive search for topological materials using symmetry
  indicators}.
\newblock {\em Nature} {\bf 2019}, {\em 566},~486.

\bibitem[Vergniory \em{et~al.}(2019)Vergniory, Elcoro, Felser, Regnault,
  Bernevig, and Wang]{vergniory2019complete}
Vergniory, M.; Elcoro, L.; Felser, C.; Regnault, N.; Bernevig, B.A.; Wang, Z.
\newblock {A complete catalogue of high-quality topological materials}.
\newblock {\em Nature} {\bf 2019}, {\em 566},~480.

\bibitem[Zheng and Jia({2019})]{zhengjia2019}
Zheng, H.; Jia, J.F.
\newblock {Topological superconductivity in a Bi2Te3/NbSe2 heterostructure: A
  review}.
\newblock {\em {\mbox{Chin. Phys. B}}} {\bf {2019}}, {\em {28}}, 067403.

\bibitem[Cr{\'e}pieux and Bruno(2001)]{crepieux2001}
Cr{\'e}pieux, A.; Bruno, P.
\newblock {Theory of the anomalous Hall effect from the Kubo formula and the
  Dirac equation}.
\newblock {\em \mbox{Phys. Rev. B}} {\bf 2001}, {\em 64},~014416.

\bibitem[Nolting(2008)]{nolting2008}
Nolting, W.
\newblock {\em {Fundamentals of Many-Body Physics}}; Springer:  Berlin/Heidelberg, Germany,
2008.

\bibitem[Ashcroft and Mermin(1976)]{ashcroft1976}
Ashcroft, N.W.; Mermin, N.D.
\newblock {\em {Solid State Physics (Saunders College, Philadelphia, 1976)}};
  Saunders College Publishing: Saunders College, PA, USA, 1976.

\bibitem[Xiao \em{et~al.}(2010)Xiao, Chang, and Niu]{Xiao2010}
Xiao, D.; Chang, M.C.; Niu, Q.
\newblock {Berry phase effects on electronic properties}.
\newblock {\em Rev. Mod. Phys.} {\bf 2010}, {\em 82},~1959--2007.

\bibitem[Nagaosa \em{et~al.}(2010)Nagaosa, Sinova, Onoda, MacDonald, and
  Ong]{nagaosa2010}
Nagaosa, N.; Sinova, J.; Onoda, S.; MacDonald, A.H.; Ong, N.P.
\newblock AnomalousHalleffe.
\newblock {\em Rev. Mod. Phys.} {\bf 2010}, {\em 82},~1539.

\bibitem[Hall(1879)]{hall1879}
Hall, E.
\newblock {On a new action of the magnet on electric currents}.
\newblock {\em Am. J. Sci.} {\bf 1879}, {\em 2},~287--292.

\bibitem[Karplus and Luttinger(1954)]{karplus1954}
Karplus, R.; Luttinger, J.
\newblock {Hall effect in ferromagnetics}.
\newblock {\em Phys. Rev.} {\bf 1954}, {\em 95},~1154.

\bibitem[Chang and Niu(1996)]{chang1996}
Chang, M.C.; Niu, Q.
\newblock {Berry phase, hyperorbits, and the Hofstadter spectrum: Semiclassical
  dynamics in magnetic Bloch bands}.
\newblock {\em Phys. Rev. B} {\bf 1996}, {\em 53},~7010.

\bibitem[Haldane(2004)]{haldane2004}
Haldane, F.
\newblock {Berry curvature on the Fermi surface: anomalous Hall effect as a
  topological Fermi-liquid property}.
\newblock {\em Phys. Rev. Lett.} {\bf 2004}, {\em 93},~206602.

\bibitem[Haldane(1988)]{Haldane1988}
Haldane, F.D.M.
\newblock Model for a Quantum Hall Eff'ect without Landau
  Levels:Condensed-Matter Realization of the ``Parity Anomaly''.
\newblock {\em Phys. Rev. Lett.} {\bf 1988}, {\em 61},~2015.

\bibitem[Halperin(1982)]{Halperin1982}
Halperin, B.I.
\newblock Quantized Hall conductance, current-carrying edge states, and the
  existence of extended states in a two-dimensional disordered potential.
\newblock {\em Phys. Rev. B} {\bf 1982}, {\em 25},~2185.

\bibitem[Yu \em{et~al.}(2010)Yu, Zhang, Zhang, Zhang, Dai, and Fang]{Yu2010}
Yu, R.; Zhang, W.; Zhang, H.J.; Zhang, S.C.; Dai, X.; Fang, Z.
\newblock Quantized Anomalous Hall Effect in Magnetic Topological Insulators.
\newblock {\em Science} {\bf 2010}, {\em 329},~61.

\bibitem[Chang \em{et~al.}(2013)Chang, Zhang, Feng, Shen, Zhang, Guo, Kang, Ou,
  Wei, Wang, Ji, Feng, Ji, Chen, , Jia, Dai, Fang, Zhang, He, Wang, Lu, Ma, and
  Xue]{Chang2013}
Chang, C.Z.; Zhang, J.; Feng, X.; Shen, J.; Zhang, Z.; Guo, M.; Kang, L.; Ou,
  Y.; Wei, P.W.; Wang, L.L.; et~al.
\newblock Experimental Observation of the Quantum Anomalous Hall Effect in a
  Magnetic Topological Insulator.
\newblock {\em Science} {\bf 2013}, {\em 340},~167.

\bibitem[Liu \em{et~al.}(2016)Liu, Zhang, and Qi]{liu2016}
Liu, C.X.; Zhang, S.C.; Qi, X.L.
\newblock {The quantum anomalous Hall effect: Theory and experiment}.
\newblock {\em Annu. Rev. Condens. Matter Phys.} {\bf 2016}, {\em
  7},~301--321.

\bibitem[Burkov \em{et~al.}(2011)Burkov, Hook, and Balents]{burkov2011}
Burkov, A.A.; Hook, M.D.; Balents, L.
\newblock {Topological nodal semimetals}.
\newblock {\em Phys. Rev. B} {\bf 2011}, {\em 84},~235126.

\bibitem[Zyuzin \em{et~al.}(2012)Zyuzin, Wu, and Burkov]{zyuzin2012}
Zyuzin, A.; Wu, S.; Burkov, A.
\newblock {Weyl semimetal with broken time reversal and inversion symmetries}.
\newblock {\em \mbox{Phys. Rev. B}} {\bf 2012}, {\em 85},~165110.

\bibitem[Lu \em{et~al.}(2015)Lu, Zhang, and Shen]{Luhaizhou2015}
Lu, H.Z.; Zhang, S.B.Z.; Shen, S.Q.
\newblock High-field magnetoconductivity of topological semimetals with
  short-range potential.
\newblock {\em Phy. Rev. B} {\bf 2015}, {\em 92},~045203.

\bibitem[K{\"u}bler and Felser(2012)]{kubler2012}
K{\"u}bler, J.; Felser, C.
\newblock {Berry curvature and the anomalous Hall effect in Heusler compounds}.
\newblock {\em Phys. Rev. B} {\bf 2012}, {\em 85},~012405.

\bibitem[Manna \em{et~al.}(2018)Manna, Muechler, Kao, Stinshoff, Zhang, Gooth,
  Kumar, Kreiner, Koepernik, Car, et~al.]{manna2018}
Manna, K.; Muechler, L.; Kao, T.H.; Stinshoff, R.; Zhang, Y.; Gooth, J.; Kumar,
  N.; Kreiner, G.; Koepernik, K.; Car, R.; et al.
\newblock {From colossal to zero: controlling the anomalous Hall effect in
  magnetic Heusler compounds via Berry curvature design}.
\newblock {\em Phys. Rev. X} {\bf 2018}, {\em 8},~041045.

\bibitem[Wang \em{et~al.}(2018)Wang, Xu, Lou, Liu, Li, Huang, Shen, Weng, Wang,
  and Lei]{wang2018}
Wang, Q.; Xu, Y.; Lou, R.; Liu, Z.; Li, M.; Huang, Y.; Shen, D.; Weng, H.;
  Wang, S.; Lei, H.
\newblock {Large intrinsic anomalous Hall effect in half-metallic ferromagnet
  Co3Sn2S2 with magnetic Weyl fermions}.
\newblock {\em Nat. Commun.} {\bf 2018}, {\em 9},~3681.

\bibitem[Liu \em{et~al.}(2018)Liu, Sun, Kumar, Muechler, Sun, Jiao, Yang, Liu,
  Liang, Xu, et~al.]{liu2018}
Liu, E.; Sun, Y.; Kumar, N.; Muechler, L.; Sun, A.; Jiao, L.; Yang, S.Y.; Liu,
  D.; Liang, A.; Xu, Q.; et al.
\newblock {Giant anomalous Hall effect in a ferromagnetic kagome-lattice
  semimetal}.
\newblock {\em Nat. Phys.} {\bf 2018}, {\em 14},~1125.

\bibitem[Kim \em{et~al.}(2018)Kim, Seo, Lee, Ko, Kim, Jang, Ok, Lee, Jo, Kang,
  et~al.]{kim2018}
Kim, K.; Seo, J.; Lee, E.; Ko, K.T.; Kim, B.; Jang, B.G.; Ok, J.M.; Lee, J.;
  Jo, Y.J.; Kang, W.; et al.
\newblock {Large anomalous Hall current induced by topological nodal lines in a
  ferromagnetic van der Waals semimetal}.
\newblock {\em Nat. Mater.} {\bf 2018}, {\em 17},~794.

\bibitem[Shindou and Nagaosa(2001)]{shindou2001}
Shindou, R.; Nagaosa, N.
\newblock {Orbital ferromagnetism and anomalous Hall effect in antiferromagnets
  on the distorted fcc lattice}.
\newblock {\em Phys. Rev. Lett.} {\bf 2001}, {\em 87},~116801.

\bibitem[Chen \em{et~al.}(2014)Chen, Niu, and MacDonald]{chen2014}
Chen, H.; Niu, Q.; MacDonald, A.H.
\newblock {Anomalous Hall effect arising from noncollinear antiferromagnetism}.
\newblock {\em Phys. Rev. Lett.} {\bf 2014}, {\em 112},~017205.

\bibitem[K{\"u}bler and Felser(2014)]{kubler2014}
K{\"u}bler, J.; Felser, C.
\newblock {Non-collinear antiferromagnets and the anomalous Hall effect}.
\newblock {\em EPL (Europhys. Lett.)} {\bf 2014}, {\em 108},~67001.

\bibitem[Nakatsuji \em{et~al.}(2015)Nakatsuji, Kiyohara, and
  Higo]{nakatsuji2015}
Nakatsuji, S.; Kiyohara, N.; Higo, T.
\newblock {Large anomalous Hall effect in a non-collinear antiferromagnet at
  room temperature}.
\newblock {\em Nature} {\bf 2015}, {\em 527},~212.

\bibitem[Nayak \em{et~al.}(2016)Nayak, Fischer, Sun, Yan, Karel, Komarek,
  Shekhar, Kumar, Schnelle, Ku~bler, Felser, and Parkin]{nayak2016}
Nayak, A.K.; Fischer, J.E.; Sun, Y.; Yan, B.; Karel, J.; Komarek, A.C.;
  Shekhar, C.; Kumar, N.; Schnelle, W.; Ku~bler, J.; et al.
\newblock {Large anomalous Hall effect driven by a nonvanishing Berry curvature
  in the noncolinear antiferromagnet Mn3Ge}.
\newblock {\em Sci. Adv.} {\bf 2016}, {\em 2},~e1501870.

\bibitem[Zhang \em{et~al.}(2017)Zhang, Sun, Yang, {\v{Z}}elezn{\`y}, Parkin,
  Felser, and Yan]{zhang2017}
Zhang, Y.; Sun, Y.; Yang, H.; {\v{Z}}elezn{\`y}, J.; Parkin, S.P.; Felser, C.;
  Yan, B.
\newblock {Strong anisotropic anomalous Hall effect and spin Hall effect in the
  chiral antiferromagnetic compounds Mn3X (X= Ge, Sn, Ga, Ir, Rh, and Pt)}.
\newblock {\em Phys. Rev. B} {\bf 2017}, {\em 95},~075128.

\bibitem[Yang \em{et~al.}(2017)Yang, Sun, Zhang, Shi, Parkin, and
  Yan]{yang2017}
Yang, H.; Sun, Y.; Zhang, Y.; Shi, W.J.; Parkin, S.S.; Yan, B.
\newblock {Topological Weyl semimetals in the chiral antiferromagnetic
  materials Mn3Ge and Mn3Sn}.
\newblock {\em New J. Phys.} {\bf 2017}, {\em 19},~015008.

\bibitem[Kuroda \em{et~al.}(2017)Kuroda, Tomita, Suzuki, Bareille, Nugroho,
  Goswami, Ochi, Ikhlas, Nakayama, Akebi, et~al.]{kuroda2017evidence}
Kuroda, K.; Tomita, T.; Suzuki, M.T.; Bareille, C.; Nugroho, A.; Goswami, P.;
  Ochi, M.; Ikhlas, M.; Nakayama, M.; Akebi, S.; et al.
\newblock {Evidence for magnetic Weyl fermions in a correlated metal}.
\newblock {\em Nat. Mater.} {\bf 2017}, {\em 16},~1090.

\bibitem[Shi \em{et~al.}(2018)Shi, Muechler, Manna, Zhang, Koepernik, Car, Van
  Den~Brink, Felser, and Sun]{shi2018prediction}
Shi, W.; Muechler, L.; Manna, K.; Zhang, Y.; Koepernik, K.; Car, R.; Van
  Den~Brink, J.; Felser, C.; Sun, Y.
\newblock {Prediction of a magnetic Weyl semimetal without spin-orbit coupling
  and strong anomalous Hall effect in the Heusler compensated ferrimagnet
  Ti2MnAl}.
\newblock {\em Phys. Rev. B} {\bf 2018}, {\em 97},~060406.

\bibitem[Markou \em{et~al.}(2019)Markou, Kriegner, Gayles, Zhang, Chen, Ernst,
  Lai, Schnelle, Chu, Sun, and Felser]{PhysRevB.100.054422}
Markou, A.; Kriegner, D.; Gayles, J.; Zhang, L.; Chen, Y.C.; Ernst, B.; Lai,
  Y.H.; Schnelle, W.; Chu, Y.H.; Sun,~Y.; et al.
\newblock Thickness dependence of the anomalous Hall effect in thin films of
  the topological semimetal ${\mathrm{Co}}_{2}\mathrm{MnGa}$.
\newblock {\em Phys. Rev. B} {\bf 2019}, {\em 100},~054422.
\newblock
  doi:{\href{https://doi.org/10.1103/PhysRevB.100.054422}{\detokenize{10.1103/PhysRevB.100.054422}}}.

\bibitem[Dulal \em{et~al.}(2019)Dulal, Dahal, Forbes, Bhattarai, Pegg, and
  Philip]{rajendra}
Dulal, R.P.; Dahal, B.R.; Forbes, A.; Bhattarai, N.; Pegg, I.L.; Philip, J.
\newblock Weak localization and small anomalous Hall conductivity in
  ferromagnetic Weyl semimetal ${\mathrm{Co}}_{2}\mathrm{TiGe}$.
\newblock {\em Sci. Rep.} {\bf 2019}, {\em 9},~3342.
\newblock
  doi:{\href{https://doi.org/10.1038/s41598-019-39037-0}{\detokenize{10.1038/s41598-019-39037-0}}}.

\bibitem[Ernst \em{et~al.}(2019)Ernst, Sahoo, Sun, Nayak, M\"uchler, Nayak,
  Kumar, Gayles, Markou, Fecher, and Felser]{PhysRevB.100.054445}
Ernst, B.; Sahoo, R.; Sun, Y.; Nayak, J.; M\"uchler, L.; Nayak, A.K.; Kumar,
  N.; Gayles, J.; Markou, A.; Fecher,~G.H.; et al.
\newblock Anomalous Hall effect and the role of Berry curvature in
  ${\mathrm{Co}}_{2}\mathrm{TiSn}$ Heusler films.
\newblock {\em Phys. Rev. B} {\bf 2019}, {\em 100},~054445.
\newblock
  doi:{\href{https://doi.org/10.1103/PhysRevB.100.054445}{\detokenize{10.1103/PhysRevB.100.054445}}}.

\bibitem[Xu \em{et~al.}(2011)Xu, Weng, Wang, Dai, and
  Fang]{PhysRevLett.107.186806}
Xu, G.; Weng, H.; Wang, Z.; Dai, X.; Fang, Z.
\newblock Chern Semimetal and the Quantized Anomalous Hall Effect in
  ${\mathrm{HgCr}}_{2}{\mathrm{Se}}_{4}$.
\newblock {\em Phys. Rev. Lett.} {\bf 2011}, {\em 107},~186806.
\newblock
  doi:{\href{https://doi.org/10.1103/PhysRevLett.107.186806}{\detokenize{10.1103/PhysRevLett.107.186806}}}.

\bibitem[Liu \em{et~al.}(2017)Liu, Hankiewicz, and Culcer]{PhysRevB.96.045307}
Liu, W.E.; Hankiewicz, E.M.; Culcer, D.
\newblock Quantum transport in Weyl semimetal thin films in the presence of
  spin-orbit coupled impurities.
\newblock {\em Phys. Rev. B} {\bf 2017}, {\em 96},~045307.
\newblock
  doi:{\href{https://doi.org/10.1103/PhysRevB.96.045307}{\detokenize{10.1103/PhysRevB.96.045307}}}.

\bibitem[Muechler \em{et~al.}(2017)Muechler, Liu, Xu, Felser, and
  Sun]{muechler2017realization}
Muechler, L.; Liu, E.; Xu, Q.; Felser, C.; Sun, Y.
\newblock Realization of quantum anomalous Hall effect from a magnetic Weyl
  semimetal.  \emph{arXiv} \textbf{2017},   	arXiv:1712.08115.

\bibitem[Bauer \em{et~al.}(2012)Bauer, Saitoh, and Van~Wees]{bauer2012spin}
Bauer, G.E.; Saitoh, E.; Van~Wees, B.J.
\newblock {Spin caloritronics}.
\newblock {\em Nat. Mater.} {\bf 2012}, {\em 11},~391.

\bibitem[Lee \em{et~al.}(2004)Lee, Watauchi, Miller, Cava, and
  Ong]{lee2004anomalous}
Lee, W.L.; Watauchi, S.; Miller, V.; Cava, R.J.; Ong, N.P.
\newblock {Anomalous Hall Heat Current and Nernst Effect in the CuCr2Se4-xBrx
  Ferromagnet}.
\newblock {\em Phys. Rev. Lett.} {\bf 2004}, {\em 93},~226601.

\bibitem[Xiao \em{et~al.}(2006)Xiao, Yao, Fang, and Niu]{xiao2006berry}
Xiao, D.; Yao, Y.; Fang, Z.; Niu, Q.
\newblock {Berry-phase effect in anomalous thermoelectric transport}.
\newblock {\em Phys. Rev. Lett.} {\bf 2006}, {\em 97},~026603.

\bibitem[Sakai \em{et~al.}(2018)Sakai, Mizuta, Nugroho, Sihombing, Koretsune,
  Suzuki, Takemori, Ishii, Nishio-Hamane, Arita, et~al.]{sakai2018giant}
Sakai, A.; Mizuta, Y.P.; Nugroho, A.A.; Sihombing, R.; Koretsune, T.; Suzuki,
  M.T.; Takemori, N.; Ishii,~R.; Nishio-Hamane, D.; Arita, R.; et al.
\newblock {Giant anomalous Nernst effect and quantum-critical scaling in a
  ferromagnetic semimetal}.
\newblock {\em Nat. Phys.} {\bf 2018}, {\em 14},~1119.

\bibitem[Guin \em{et~al.}(2019{\natexlab{a}})Guin, Manna, Noky, Watzman, Fu,
  Kumar, Schnelle, Shekhar, Sun, Gooth, et~al.]{guin2019anomalous}
Guin, S.N.; Manna, K.; Noky, J.; Watzman, S.J.; Fu, C.; Kumar, N.; Schnelle,
  W.; Shekhar, C.; Sun, Y.; Gooth,~J.; et al.
\newblock {Anomalous Nernst effect beyond the magnetization scaling relation in
  the ferromagnetic Heusler compound Co2MnGa}.
\newblock {\em NPG Asia Mater.} {\bf 2019}, {\em 11},~16.

\bibitem[Guin \em{et~al.}(2019{\natexlab{b}})Guin, Vir, Zhang, Kumar, Watzman,
  Fu, Liu, Manna, Schnelle, Gooth, et~al.]{guin2019zero}
Guin, S.N.; Vir, P.; Zhang, Y.; Kumar, N.; Watzman, S.J.; Fu, C.; Liu, E.;
  Manna, K.; Schnelle, W.; Gooth, J.; et al.
\newblock {Zero-Field Nernst Effect in a Ferromagnetic Kagome-Lattice
  Weyl-Semimetal Co3Sn2S2}.
\newblock {\em Adv. Mater.} {\bf 2019}, \emph{2019}, 1806622.

\bibitem[Hirokane \em{et~al.}({2016})Hirokane, Tomioka, Imai, Maeda, and
  Onose]{Hirokane2016}
Hirokane, Y.; Tomioka, Y.; Imai, Y.; Maeda, A.; Onose, Y.
\newblock {Longitudinal and transverse thermoelectric transport in MnSi}.
\newblock {\em {Phys. Rev. B}} {\bf {2016}}, {\em {93}}.
\newblock
  doi:{\href{https://doi.org/{10.1103/PhysRevB.93.014436}}{\detokenize{10.1103/PhysRevB.93.014436}}}.

\bibitem[Miyasato \em{et~al.}({2007})Miyasato, Abe, Fujii, Asamitsu, Onoda,
  Onose, Nagaosa, and Tokura]{Miyasato2007}
Miyasato, T.; Abe, N.; Fujii, T.; Asamitsu, A.; Onoda, S.; Onose, Y.; Nagaosa,
  N.; Tokura, Y.
\newblock {Crossover behavior of the anomalous hall effect and anomalous nernst
  effect in itinerant ferromagnets}.
\newblock {\em {Phys. Rev. Lett.}} {\bf {2007}}, {\em {99}}.
\newblock
  doi:{\href{https://doi.org/{10.1103/PhysRevLett.99.086602}}{\detokenize{10.1103/PhysRevLett.99.086602}}}.

\bibitem[Ikhlas \em{et~al.}({2017})Ikhlas, Tomita, Koretsune, Suzuki,
  Nishio-Hamane, Arita, Otani, and Nakatsuji]{Ikhlas2017}
Ikhlas, M.; Tomita, T.; Koretsune, T.; Suzuki, M.T.; Nishio-Hamane, D.; Arita,
  R.; Otani, Y.; Nakatsuji, S.
\newblock {Large anomalous Nernst effect at room temperature in a chiral
  antiferromagnet}.
\newblock {\em {Nat. Phys.}} {\bf {2017}}, {\em {13}},~{1085}.
\newblock
  doi:{\href{https://doi.org/{10.1038/NPHYS4181}}{\detokenize{10.1038/NPHYS4181}}}.

\bibitem[Li \em{et~al.}({2017})Li, Xu, Ding, Wang, Shen, Lu, Zhu, and
  Behnia]{Li2017}
Li, X.; Xu, L.; Ding, L.; Wang, J.; Shen, M.; Lu, X.; Zhu, Z.; Behnia, K.
\newblock {Anomalous Nernst and Righi-Leduc Effects in Mn3Sn: Berry Curvature
  and Entropy Flow}.
\newblock {\em {Phys. Rev. Lett.}} {\bf {2017}}, {\em {119}}.
\newblock
  doi:{\href{https://doi.org/{10.1103/PhysRevLett.119.056601}}{\detokenize{10.1103/PhysRevLett.119.056601}}}.

\bibitem[Hanasaki \em{et~al.}({2008})Hanasaki, Sano, Onose, Ohtsuka, Iguchi,
  Kezsmarki, Miyasaka, Onoda, Nagaosa, and Tokura]{Hanasaki2008}
Hanasaki, N.; Sano, K.; Onose, Y.; Ohtsuka, T.; Iguchi, S.; Kezsmarki, I.;
  Miyasaka, S.; Onoda, S.; Nagaosa,~N.; Tokura, Y.
\newblock {Anomalous nernst effects in pyrochlore molybdates with spin
  chirality}.
\newblock {\em {Phys. Rev. Lett.}} {\bf {2008}}, {\em {100}}.
\newblock
  doi:{\href{https://doi.org/{10.1103/PhysRevLett.100.106601}}{\detokenize{10.1103/PhysRevLett.100.106601}}}.

\bibitem[Pu \em{et~al.}({2008})Pu, Chiba, Matsukura, Ohno, and Shi]{Pu2008}
Pu, Y.; Chiba, D.; Matsukura, F.; Ohno, H.; Shi, J.
\newblock {Mott relation for anomalous Hall and Nernst effects in Ga1-xMnxAs
  ferromagnetic semiconductors}.
\newblock {\em {Phys. Rev. Lett.}} {\bf {2008}}, {\em {101}}.
\newblock
  doi:{\href{https://doi.org/{10.1103/PhysRevLett.101.117208}}{\detokenize{10.1103/PhysRevLett.101.117208}}}.

\bibitem[Ramos \em{et~al.}({2014})Ramos, Aguirre, Anadon, Blasco, Lucas,
  Uchida, Algarabel, Morellon, Saitoh, and Ibarra]{Ramos2014}
Ramos, R.; Aguirre, M.H.; Anadon, A.; Blasco, J.; Lucas, I.; Uchida, K.;
  Algarabel, P.A.; Morellon, L.; Saitoh, E.; Ibarra, M.R.
\newblock {Anomalous Nernst effect of Fe3O4 single crystal}.
\newblock {\em {Phys. Rev. B}} {\bf {2014}}, {\em {90}}.
\newblock
  doi:{\href{https://doi.org/{10.1103/PhysRevB.90.054422}}{\detokenize{10.1103/PhysRevB.90.054422}}}.

\bibitem[Weischenberg \em{et~al.}({2013})Weischenberg, Freimuth, Bluegel, and
  Mokrousov]{Weischenberg2013}
Weischenberg, J.; Freimuth, F.; Bluegel, S.; Mokrousov, Y.
\newblock {Scattering-independent anomalous Nernst effect in ferromagnets}.
\newblock {\em {Phys. Rev. B}} {\bf {2013}}, {\em {87}}.
\newblock
  doi:{\href{https://doi.org/{10.1103/PhysRevB.87.060406}}{\detokenize{10.1103/PhysRevB.87.060406}}}.

\bibitem[Noky \em{et~al.}(2018)Noky, Gooth, Felser, and
  Sun]{noky2018characterization}
Noky, J.; Gooth, J.; Felser, C.; Sun, Y.
\newblock {Characterization of topological band structures away from the Fermi
  level by the anomalous Nernst effect}.
\newblock {\em Phys. Rev. B} {\bf 2018}, {\em 98},~241106.

\bibitem[D'yakonov and Perel(1971)]{d1971possibility}
D'yakonov, M.; Perel, V.
\newblock {Possibility of orienting electron spins with current}.
\newblock {\em Sov. J. Exp. Theor. Phys. Lett.}
  {\bf 1971}, {\em 13},~467.

\bibitem[Hirsch(1999)]{hirsch1999spin}
Hirsch, J.
\newblock {Spin hall effect}.
\newblock {\em Phys. Rev. Lett.} {\bf 1999}, {\em 83},~1834.

\bibitem[Kato \em{et~al.}(2004)Kato, Myers, Gossard, and
  Awschalom]{kato2004observation}
Kato, Y.K.; Myers, R.C.; Gossard, A.C.; Awschalom, D.D.
\newblock {Observation of the spin Hall effect in semiconductors}.
\newblock {\em Science} {\bf 2004}, {\em 306},~1910--1913.

\bibitem[Sinova \em{et~al.}(2015)Sinova, Valenzuela, Wunderlich, Back, and
  Jungwirth]{sinova2015spin}
Sinova, J.; Valenzuela, S.O.; Wunderlich, J.; Back, C.; Jungwirth, T.
\newblock {Spin hall effects}.
\newblock {\em Rev. Mod. Phys.} {\bf 2015}, {\em 87},~1213.

\bibitem[Murakami \em{et~al.}(2003)Murakami, Nagaosa, and Zhang]{Murakami2003}
Murakami, S.; Nagaosa, N.; Zhang, S.C.
\newblock Dissipationless Quantum Spin Current at Room Temperature.
\newblock {\em Science} {\bf 2003}, {\em 301},~1348.

\bibitem[Sinova \em{et~al.}(2004)Sinova, Culcer, Niu, Sinitsyn, Jungwirth, and
  MacDonald]{sinova2004universal}
Sinova, J.; Culcer, D.; Niu, Q.; Sinitsyn, N.; Jungwirth, T.; MacDonald, A.H.
\newblock {Universal intrinsic spin Hall effect}.
\newblock {\em Phys. Rev. Lett.} {\bf 2004}, {\em 92},~126603.

\bibitem[Bernevig and Zhang(2005)]{Bernevig2005}
Bernevig, B.A.; Zhang, S.C.
\newblock Intrinsic Spin Hall Effect in the Two-Dimensional Hole Gas.
\newblock {\em Phys. Rev. Lett.} {\bf 2005}, {\em 95},~016801.

\bibitem[Wunderlich \em{et~al.}(2005)Wunderlich, Kaestner, Sinova, and
  Jungwirth]{Wunderlich2005}
Wunderlich, J.; Kaestner, B.; Sinova, J.; Jungwirth, T.
\newblock Experimental Observation of the Spin-Hall Effect in a Two-Dimensional
  Spin-Orbit Coupled Semiconductor System.
\newblock {\em Phy. Rev. Lett.} {\bf 2005}, {\em 94},~047204.

\bibitem[Day(2005)]{Day2005}
Day, C.
\newblock Two Groups Observe the Spin Hall Effect in Semiconductors.
\newblock {\em Phys. Today} {\bf 2005}, {\em 58},~17.

\bibitem[Liu \em{et~al.}({2012})Liu, Pai, Li, Tseng, Ralph, and
  Buhrman]{Liu2012}
Liu, L.; Pai, C.F.; Li, Y.; Tseng, H.W.; Ralph, D.C.; Buhrman, R.A.
\newblock {Spin-Torque Switching with the Giant Spin Hall Effect of Tantalum}.
\newblock {\em {Science}} {\bf {2012}}, {\em {336}},~{555--558}.
\newblock
  doi:{\href{https://doi.org/{10.1126/science.1218197}}{\detokenize{10.1126/science.1218197}}}.

\bibitem[Pai \em{et~al.}({2012})Pai, Liu, Li, Tseng, Ralph, and
  Buhrman]{Pai2012}
Pai, C.F.; Liu, L.; Li, Y.; Tseng, H.W.; Ralph, D.C.; Buhrman, R.A.
\newblock {Spin transfer torque devices utilizing the giant spin Hall effect of
  tungsten}.
\newblock {\em {Appl. Phys. Lett.}} {\bf {2012}}, {\em {101}}.
\newblock
  doi:{\href{https://doi.org/{10.1063/1.4753947}}{\detokenize{10.1063/1.4753947}}}.

\bibitem[Liu \em{et~al.}(2011)Liu, Lee, Kondo, Mun, Caudle, Harmon, Bud'ko,
  Canfield, and Kaminski]{Liu2011}
Liu, C.; Lee, Y.; Kondo, T.; Mun, E.D.; Caudle, M.; Harmon, B.N.; Bud'ko, S.L.;
  Canfield, P.C.; Kaminski,~A.
\newblock {Metallic surface electronic state in half-Heusler compounds $R$PtBi
  ($R$$=$ Lu, Dy, Gd)}.
\newblock {\em Phys. Rev. B} {\bf 2011}, {\em 83},~205133.

\bibitem[Mellnik \em{et~al.}(2014)Mellnik, Lee, Richardella, Grab, Mintun,
  Fischer, Vaezi, Manchon, Kim, Samarth, et~al.]{mellnik2014spin}
Mellnik, A.; Lee, J.; Richardella, A.; Grab, J.; Mintun, P.; Fischer, M.H.;
  Vaezi, A.; Manchon, A.; Kim, E.A.; Samarth, N.; et al.
\newblock {Spin-transfer torque generated by a topological insulator}.
\newblock {\em Nature} {\bf 2014}, {\em 511},~449.

\bibitem[Mahendra \em{et~al.}(2018)Mahendra, Grassi, Chen, Jamali, Hickey,
  Zhang, Zhao, Li, Quarterman, Lv, et~al.]{mahendra2018room}
Mahendra, D.; Grassi, R.; Chen, J.Y.; Jamali, M.; Hickey, D.R.; Zhang, D.;
  Zhao, Z.; Li, H.; Quarterman, P.; Lv,~Y.; et al.
\newblock {Room-temperature high spin orbit torque due to quantum confinement
  in sputtered BixSe(1-x) films}.
\newblock {\em Nat. Mater.} {\bf 2018}, {\em 17},~800.
  
\bibitem[Khang \em{et~al.}(2018)Khang, Ueda, and Hai]{khang2018conductive}
Khang, N.H.D.; Ueda, Y.; Hai, P.N.
\newblock {A conductive topological insulator with large spin Hall effect for
  ultralow power spin--orbit torque switching}.
\newblock {\em Nat. Mater.} {\bf 2018}, \emph{17}, 808–813.

\bibitem[Tanaka \em{et~al.}(2008)Tanaka, Kontani, Naito, Naito, Hirashima,
  Yamada, and Inoue]{Tanaka2008}
Tanaka, T.; Kontani, H.; Naito, M.; Naito, T.; Hirashima, D.S.; Yamada, K.;
  Inoue, J.
\newblock {Intrinsic spin Hall effect and orbital Hall effect in 4\emph{d} and
  5\emph{d} transition metals}.
\newblock {\em Phys. Rev. B} {\bf 2008}, {\em 77},~165117.

\bibitem[Hoffmann(2013)]{Hoffmann2013}
Hoffmann, A.
\newblock Advances in Magnetics: Spin Hall Effects in Metals.
\newblock {\em IEEE Trans. Magn.} {\bf 2013}, {\em 49},~5172.

\bibitem[Kimura \em{et~al.}(2007)Kimura, Otani, Sato, Takahashi, and
  Maekawa]{Kimura2007}
Kimura, T.; Otani, Y.; Sato, T.; Takahashi, S.; Maekawa, S.
\newblock Room-Temperature Reversible Spin Hall Effect.
\newblock {\em Phys. Rev. Lett.} {\bf 2007}, {\em 98},~156601.

\bibitem[Saitoh \em{et~al.}(2006)Saitoh, Ueda, Miyajima, and
  Tatara]{Saitoh2006}
Saitoh, E.; Ueda, M.; Miyajima, H.; Tatara, G.
\newblock {Conversion of spin current into charge current at room temperature:
  Inverse spin-Hall effect}.
\newblock {\em Appl. Phys. Lett.} {\bf 2006}, {\em 88},~182509.

\bibitem[Fan \em{et~al.}(2014)Fan, Upadhyaya, Kou, Lang, Takei, Wang, Tang, He,
  Chang, Montazeri, et~al.]{fan2014magnetization}
Fan, Y.; Upadhyaya, P.; Kou, X.; Lang, M.; Takei, S.; Wang, Z.; Tang, J.; He,
  L.; Chang, L.T.; Montazeri,~M.; et~al.
\newblock {Magnetization switching through giant spin--orbit torque in a
  magnetically doped topological insulator heterostructure}.
\newblock {\em Nat. Mater.} {\bf 2014}, {\em 13},~699.

\bibitem[Zhang \em{et~al.}(2019)Zhang, Xu, Koepernik, Zelezny, Jungwirth,
  Felser, Brink, and Sun]{zhangyang2019}
Zhang, Y.; Xu, Q.; Koepernik, K.; Zelezny, J.; Jungwirth, T.; Felser, C.;
  Brink, J.V.d.; Sun, Y.S.
\newblock {Spin-orbitronic materials with record spin-charge conversion from
  high-throughput ab initio calculations}.
\newblock {\em arXiv} {\bf 2019}, arXiv:1909.09605.

\bibitem[Meyer \em{et~al.}(2017)Meyer, Chen, Wimmer, Althammer, Wimmer,
  Schlitz, Geprags, Huebl, Ködderitzsch, Ebert, Bauer, Gross, and
  Goennenwein]{Meyer2017}
Meyer, S.; Chen, Y.T.; Wimmer, S.; Althammer, M.; Wimmer, T.; Schlitz, R.;
  Geprags, S.; Huebl, H.; Ködderitzsch, D.; Ebert, H.; et al.
\newblock Observation of the spin Nernst effect.
\newblock {\em Nat. Mater.} {\bf 2017}, {\em 16},~977.

\bibitem[Sheng \em{et~al.}(2017)Sheng, Sakuraba, Lau, Takahashi, Mitani, and
  Hayashi]{sheng2017spin}
Sheng, P.; Sakuraba, Y.; Lau, Y.C.; Takahashi, S.; Mitani, S.; Hayashi, M.
\newblock {The spin Nernst effect in tungsten}.
\newblock {\em Sci. Adv.} {\bf 2017}, {\em 3},~e1701503.

\bibitem[Kim \em{et~al.}(2017)Kim, Jeon, Choi, Lee, Surabhi, Jeong, Lee, and
  Park]{kim2017observation}
Kim, D.J.; Jeon, C.Y.; Choi, J.G.; Lee, J.W.; Surabhi, S.; Jeong, J.R.; Lee,
  K.J.; Park, B.G.
\newblock {Observation of transverse spin Nernst magnetoresistance induced by
  thermal spin current in ferromagnet/non-magnet bilayers}.
\newblock {\em Nat. Commun.} {\bf 2017}, {\em 8},~1400.

\bibitem[Manchon and Zhang(2009)]{manchon2009theory}
Manchon, A.; Zhang, S.
\newblock {Theory of spin torque due to spin-orbit coupling}.
\newblock {\em Phys. Rev. B} {\bf 2009}, {\em 79},~094422.

\bibitem[Garate and MacDonald(2009)]{garate2009influence}
Garate, I.; MacDonald, A.H.
\newblock {Influence of a transport current on magnetic anisotropy in
  gyrotropic ferromagnets}.
\newblock {\em Phys. Rev. B} {\bf 2009}, {\em 80},~134403.

\bibitem[Chernyshov \em{et~al.}(2009)Chernyshov, Overby, Liu, Furdyna,
  Lyanda-Geller, and Rokhinson]{chernyshov2009evidence}
Chernyshov, A.; Overby, M.; Liu, X.; Furdyna, J.K.; Lyanda-Geller, Y.;
  Rokhinson, L.P.
\newblock {Evidence for reversible control of magnetization in a ferromagnetic
  material by means of spin orbit magnetic field}.
\newblock {\em Nat. Phys.} {\bf 2009}, {\em 5},~656.

\bibitem[Kurebayashi \em{et~al.}(2014)Kurebayashi, Sinova, Fang, Irvine,
  Skinner, Wunderlich, Nov{\'a}k, Campion, Gallagher, Vehstedt,
  et~al.]{kurebayashi2014antidamping}
Kurebayashi, H.; Sinova, J.; Fang, D.; Irvine, A.; Skinner, T.; Wunderlich, J.;
  Nov{\'a}k, V.; Campion, R.; Gallagher, B.; Vehstedt, E.; et al.
\newblock {An antidamping spin orbit torque originating from the Berry
  curvature}.
\newblock {\em Nat. Nanotechnol.} {\bf 2014}, {\em 9},~211.

\bibitem[Miron \em{et~al.}(2010)Miron, Gaudin, Auffret, Rodmacq, Schuhl,
  Pizzini, Vogel, and Gambardella]{miron2010current}
Miron, I.M.; Gaudin, G.; Auffret, S.; Rodmacq, B.; Schuhl, A.; Pizzini, S.;
  Vogel, J.; Gambardella, P.
\newblock {Current-driven spin torque induced by the Rashba effect in a
  ferromagnetic metal layer}.
\newblock {\em Nat. Mater.} {\bf 2010}, {\em 9},~230.

\bibitem[Miron \em{et~al.}(2011)Miron, Garello, Gaudin, Zermatten, Costache,
  Auffret, Bandiera, Rodmacq, Schuhl, and Gambardella]{miron2011perpendicular}
Miron, I.M.; Garello, K.; Gaudin, G.; Zermatten, P.J.; Costache, M.V.; Auffret,
  S.; Bandiera, S.; Rodmacq, B.; Schuhl, A.; Gambardella, P.
\newblock {Perpendicular switching of a single ferromagnetic layer induced by
  in-plane current injection}.
\newblock {\em Nature} {\bf 2011}, {\em 476},~189.

\bibitem[Liu \em{et~al.}(2012)Liu, Lee, Gudmundsen, Ralph, and
  Buhrman]{liu2012current}
Liu, L.; Lee, O.; Gudmundsen, T.; Ralph, D.; Buhrman, R.
\newblock {Current-induced switching of perpendicularly magnetized magnetic
  layers using spin torque from the spin Hall effect}.
\newblock {\em Phys. Rev. Lett.} {\bf 2012}, {\em 109},~096602.

\bibitem[Pi \em{et~al.}({2010})Pi, Kim, Bae, Lee, Cho, Kim, and Seo]{Pi2010}
Pi, U.H.; Kim, K.W.; Bae, J.Y.; Lee, S.C.; Cho, Y.J.; Kim, K.S.; Seo, S.
\newblock {Tilting of the spin orientation induced by Rashba effect in
  ferromagnetic metal layer}.
\newblock {\em {Appl. Phys. Lett.}} {\bf {2010}}, {\em {97}}.
\newblock
  doi:{\href{https://doi.org/{10.1063/1.3502596}}{\detokenize{10.1063/1.3502596}}}.

\bibitem[Kim \em{et~al.}({2013})Kim, Sinha, Hayashi, Yamanouchi, Fukami,
  Suzuki, Mitani, and Ohno]{KimJunyoen2013}
Kim, J.; Sinha, J.; Hayashi, M.; Yamanouchi, M.; Fukami, S.; Suzuki, T.;
  Mitani, S.; Ohno, H.
\newblock {Layer thickness dependence of the current-induced effective field
  vector in Ta vertical bar CoFeB vertical bar MgO}.
\newblock {\em {Nat.~Mater.}} {\bf {2013}}, {\em {12}},~{240--245}.
\newblock
  doi:{\href{https://doi.org/{10.1038/NMAT3522}}{\detokenize{10.1038/NMAT3522}}}.

\bibitem[Suzuki \em{et~al.}({2011})Suzuki, Fukami, Ishiwata, Yamanouchi, Ikeda,
  Kasai, and Ohno]{Suzuki2011}
Suzuki, T.; Fukami, S.; Ishiwata, N.; Yamanouchi, M.; Ikeda, S.; Kasai, N.;
  Ohno, H.
\newblock {Current-induced effective field in perpendicularly magnetized
  Ta/CoFeB/MgO wire}.
\newblock {\em {Appl. Phys. Lett.}} {\bf {2011}}, {\em {98}}.
\newblock
  doi:{\href{https://doi.org/{10.1063/1.3579155}}{\detokenize{10.1063/1.3579155}}}.

\bibitem[Gomonay and Loktev({2014})]{Gomonay2014}
Gomonay, E.V.; Loktev, V.M.
\newblock {Spintronics of antiferromagnetic systems}.
\newblock {\em {Low Temp. Phys.}} {\bf {2014}}, {\em {40}},~{17--35}.
\newblock
  doi:{\href{https://doi.org/{10.1063/1.4862467}}{\detokenize{10.1063/1.4862467}}}.

\bibitem[MacDonald and Tsoi({2011})]{MacDonald2011}
MacDonald, A.H.; Tsoi, M.
\newblock {Antiferromagnetic metal spintronics}.
\newblock {\em {Philos. Trans. R. Soc. A Math. Phys. Eng. Sci.}} {\bf {2011}}, {\em {369}},~{3098--3114}.
\newblock
  doi:{\href{https://doi.org/{10.1098/rsta.2011.0014}}{\detokenize{10.1098/rsta.2011.0014}}}.

\bibitem[Jungwirth \em{et~al.}({2016})Jungwirth, Marti, Wadley, and
  Wunderlich]{Jungwirth2016}
Jungwirth, T.; Marti, X.; Wadley, P.; Wunderlich, J.
\newblock {Antiferromagnetic spintronics}.
\newblock {\em {Nat. Nanotechnol.}} {\bf {2016}}, {\em {11}},~{231--241}.
\newblock
  doi:{\href{https://doi.org/{10.1038/NNANO.2016.18}}{\detokenize{10.1038/NNANO.2016.18}}}.

\bibitem[Zelezny \em{et~al.}({2014})Zelezny, Gao, Vyborny, Zemen, Masek,
  Manchon, Wunderlich, Sinova, and Jungwirth]{Zelezny2014}
Zelezny, J.; Gao, H.; Vyborny, K.; Zemen, J.; Masek, J.; Manchon, A.;
  Wunderlich, J.; Sinova, J.; Jungwirth, T.
\newblock {Relativistic Neel-Order Fields Induced by Electrical Current in
  Antiferromagnets}.
\newblock {\em {Phys. Rev. Lett.}} {\bf {2014}}, {\em {113}}.
\newblock
  doi:{\href{https://doi.org/{10.1103/PhysRevLett.113.157201}}{\detokenize{10.1103/PhysRevLett.113.157201}}}.

\bibitem[Wadley \em{et~al.}({2016})Wadley, Howells, Zelezny, Andrews, Hills,
  Campion, Novak, Olejnik, Maccherozzi, Dhesi, Martin, Wagner, Wunderlich,
  Freimuth, Mokrousov, Kunes, Chauhan, Grzybowski, Rushforth, Edmonds,
  Gallagher, and Jungwirth]{Wadley2016}
Wadley, P.; Howells, B.; Zelezny, J.; Andrews, C.; Hills, V.; Campion, R.P.;
  Novak, V.; Olejnik, K.; Maccherozzi, F.; Dhesi, S.S.; et al.
\newblock {Electrical switching of an antiferromagnet}.
\newblock {\em {Science}} {\bf {2016}}, {\em {351}},~{587--590}.
\newblock
  doi:{\href{https://doi.org/{10.1126/science.aab1031}}{\detokenize{10.1126/science.aab1031}}}.

\bibitem[Tang \em{et~al.}(2016)Tang, Zhou, Xu, and Zhang]{Tang2016}
Tang, P.; Zhou, Q.; Xu, G.; Zhang, S.C.
\newblock {Dirac fermions in an antiferromagnetic semimetal}.
\newblock {\em Nat. Phys.} {\bf 2016}, {doi:10.1038/nphys3839}.

\bibitem[Smejkal \em{et~al.}({2017})Smejkal, Zelezny, Sinova, and
  Jungwirth]{Smejkal2017}
Smejkal, L.; Zelezny, J.; Sinova, J.; Jungwirth, T.
\newblock {Electric Control of Dirac Quasiparticles by Spin-Orbit Torque in an
  Antiferromagnet}.
\newblock {\em {Phys. Rev. Lett.}} {\bf {2017}}, {\em {118}}.
\newblock
  doi:{\href{https://doi.org/{10.1103/PhysRevLett.118.106402}}{\detokenize{10.1103/PhysRevLett.118.106402}}}.

\end{thebibliography}
\end{document}